\documentclass[letterpaper, 11pt]{article}

\pdfoutput=1

\usepackage{amsmath, amssymb, amsfonts}
\usepackage{bm, bbm}
\usepackage{epsfig}
\usepackage{graphicx}

\usepackage[in, plain]{fullpage}

\usepackage{slashed}
\usepackage{color}
\usepackage{comment}
\usepackage{hyperref}
\usepackage{caption}
\usepackage{subcaption}

\hypersetup{colorlinks,bookmarksopen,bookmarksnumbered,citecolor=blue,
	pdfstartview=FitH}

\newcommand\beq{\begin{equation}}
\newcommand\eeq{\end{equation}}

\newcommand\nn{\nonumber}

\newcommand{\PD}{{\ensuremath{\phantom{\dagger}}}} 
\newcommand\hc{\text{h.c.}}

\newcommand{\fref}[1]{Fig.~\ref{f.#1}}
\newcommand{\eref}[1]{Eq.~(\ref{e.#1})}
\newcommand{\erefn}[1]{(\ref{e.#1})}
\newcommand{\erefs}[2]{Eqs.~(\ref{e.#1})--(\ref{e.#2})}

\newcommand{\sref}[1]{Section~\ref{s.#1}}

\newcommand{\cref}[1]{Chapter~\ref{c.#1}}
\newcommand{\tref}[1]{Table~\ref{t.#1}}

\def\cl{{\mathcal L}}

\def\co{{\mathcal O}}

\newcommand{\gsu}[1]{{\mathrm{SU}(#1)}}
\newcommand{\gu}[1]{{\mathrm{U}(#1)}}

\def\fb{\, {\rm fb}}

\def\gev{\>{\rm GeV}}

\def\mev{\>{\rm MeV}}
\def\tev{\>{\rm TeV}}

\def\MET{\slashed{E}_\mathrm{T}}

\newcommand{\simlt}{\lesssim}
\newcommand{\simgt}{\gtrsim}

\newcommand{\I}{\mathrm{i}}
\newcommand{\C}{\mathrm{c}}
\newcommand{\X}{\mathrm{X}}
\newcommand{\T}{\mathrm{T}}
\newcommand{\LL}{\text{\tiny{L}}}
\newcommand{\RR}{\text{\tiny{R}}}
\newcommand{\W}{\text{\tiny{$W$}}}
\newcommand{\Z}{\text{\tiny{$Z$}}}
\newcommand{\chiL}{\raise0.15em\hbox{$\chi$}_{0\LL}}
\newcommand{\thetaW}{\theta_\text{\tiny{W}}}

\newcommand{\D}[1]{\mathrm{D}_{\hspace{-0.2ex}#1}}

\newcommand{\EQ}[1]{$\displaystyle{#1}$}

\begin{document}

\begin{titlepage}

\setcounter{page}{0}

~\\
\vskip 1in

\begin{center}

{\huge\bf Higgs Production Amidst the LHC Detector}

\vskip .5in

{\large {\bf Prerit Jaiswal,}$^1$ {\bf Karoline K\"{o}pp,}$^2$ and {\bf Takemichi Okui}$^3$}

\vskip .25in

{\it Department of Physics, Florida State University, Tallahassee, FL 32306}

\vskip 1in

\abstract{We investigate the spectacular collider signatures of macroscopically displaced, neutral particles that decay to Higgs bosons and missing energy. We show that such long-lived particles arise naturally in a very minimal extension of the Standard Model with only two new fermions with electroweak interactions. The lifetime of the long-lived neutral particles can range from $10^{-2}$~mm to $10^6$~mm. In some regions of the parameter space, the exotic signals would have already been selected by the ATLAS and CMS triggers in their $7$ and $8$ TeV runs, hence hiding in the existing data. We also discuss the possibility of explaining the mild anomalies observed in the diphoton Higgs channel and the $WW$ production at the LHC.}

\end{center}

\vfill
\begin{flushleft}
{\tt\small 
$^1$~prerit.jaiswal@hep.fsu.edu\\
$^2$~kk09t@my.fsu.edu\\
$^3$~okui@hep.fsu.edu}
\end{flushleft}

\end{titlepage}

\section{Introduction}
\label{s.Introduction}
The recent discovery of a new resonance with mass $\approx 125\gev$~\cite{Higgs125Combined} at the Large Hadron Collider (LHC) 
is an extraordinary milestone for fundamental physics. 
Although the properties of the resonance are consistent 
with those of the Higgs boson of the Standard Model (SM) at the $2\sigma$ level, 
the experimental uncertainties are still large and allow $ \sim10\,\%$ deviations from the SM\@, 
even if we put aside the potential indications of a mild excess in the diphoton channel, 
\EQ{pp \to h \to \gamma \gamma}~\cite{Higgs125gaga}.
If the diphoton excess or any other deviation in the Higgs properties is indeed confirmed, 
it means the presence of new physics beyond the SM at the electroweak scale. 

In this paper, we explore the possibility of new physics \emph{right at the 100-GeV scale}  
providing highly non-standard Higgs signals that elude existing search strategies aimed at the SM Higgs or its close variants. 
Specifically, the Higgs boson can be produced in association with missing energy 
from the decay of a new, \emph{macroscopically long-lived, neutral} particle produced via electroweak gauge interaction. 
In some regions of the parameter space, it is even possible that the signals we propose are already in the existing LHC data, 
while in other regions the near-future discovery potential is high. 
Our work differs from most of the existing literature on non-standard Higgs phenomenology, which have either focussed on exotic Higgs \emph{decays} ~\cite{HiddenValley_Higgs, NMSSM, Hidden_Higgs, DisplacedSUSY}, or \emph{prompt} Higgs production from new physics, for example,  in the context of Gauge-Mediated Supersymmetry Breaking (GMSB) ~\cite{GMSB, GMSB:Prompt}, though displaced Higgs scenario has been briefly discussed in \cite{DisplacedDM, SplitSusy}

The extension of the SM proposed in this paper contains only two new fields: 
a massless singlet fermion and a massive electroweak-triplet fermion.%
\footnote{These new states could further be embedded in a supersymmetric model if one wishes, 
although we will not go through such model building exercise in this paper.}
We do not introduce any new gauge interactions beyond the SM\@.
Our model thus significantly differs from the hidden valley models~\cite{HiddenValley_Original}, 
which involve new confining gauge interactions, 
although a class of hidden valley models can also give rise to non-standard Higgs signals similar to the ones we propose~\cite{HiddenValley_Higgs}.
The lagrangian of our model and its field theoretic properties will be discussed in detail in \sref{Model}.

Despite its minimality, our model displays a strikingly rich collider phenomenology, with a two-dimensional parameter space 
divided into eight signal regions, depending on the spectrum and dominant decay modes. 
In \sref{Pheno}, we will define these signal regions 
and analyze the constraints and discovery potential for each region.
We discuss how our model could lead to a diphoton excess. 
and also explain the mild anomaly in the \EQ{W^+ W^-} sample observed at the LHC~\cite{WW:SM} 
in a manner similar to the charginos in supersymmetric models~\cite{WWanomaly}.

\section{The Model}
\label{s.Model}
We consider a model with additional spin-$1/2$ fermions that are vectorlike under the SM gauge group. 
We introduce a Dirac fermion $\omega$ and a left-handed Weyl fermion $\chiL$ with SM gauge quantum numbers 
\EQ{(\mathbf{1}, \mathbf{3})_0} and \EQ{(\mathbf{1}, \mathbf{1})_0}, respectively. 
We assume a new unbroken global $\gu{1}$ symmetry, $\gu{1}_\X$, 
under which $\omega$ and $\chiL$ both carry charge $+1$,%
\footnote{$\gu{1}_\X$ is anomalous in the presence of gravity, but the anomaly could easily be cancelled  
by simply adding extra SM singlet fermions with $\gu{1}_\X$ charge, as there are no anomalies induced by 
SM gauge interactions.}
the importance of which will be explained shortly.
After electroweak symmetry breaking (EWSB), we will use the electric-charge basis defined as
\begin{align}
\omega_+ \equiv \frac{\omega^1 - \I \omega^2}{\sqrt2}
\,,\quad 
\omega'_- \equiv \frac{\omega^1 + \I \omega^2}{\sqrt2}
\,,\quad
\omega_0 \equiv \omega^3 
\,,
\label{e.Mbasis}
\end{align}
where $\omega^a$ ($a=1$, $2$, $3$) label the components of the $\gsu{2}_\LL$-triplet $\omega$. 
The three fermion fields $\omega_+$, $\omega'_-$ and $\omega_0$ 
carry electric charge $+1$, $-1$ and $0$, respectively, 
while they all have $\gu{1}_\X$ charge $+1$.
The antiparticle of $\omega_+$ will be referred to as $\omega_-$, 
which is distinct from $\omega'_-$, as the former carries $\gu{1}_\X$ charge $-1$ while the latter $+1$.
As we will see below, $\omega_\pm$ and $\omega'_\mp$ may even have different masses.
The antiparticle of $\omega_0$ (carrying $\gu{1}_\X$ charge $-1$) will be referred to as $\overline{\omega}_0$.
Similarly, the particle and antiparticle interpolated by the $\chiL$ field will be denoted by $\chi_0$ and $\overline{\chi}_0$, 
respectively.
The most general renormalizable lagrangian $\cl_\text{ren}$ 
consistent with SM gauge invariance and the $\gu{1}_\X$ symmetry is given by
\begin{equation}
\cl_\text{ren} = 
\I \overline{\omega} \gamma^{\mu} \D{\mu} \omega 
- m_{\omega\,} \overline{\omega} \omega 
+ \I \chiL^\dag \overline{\sigma}^\mu \partial_\mu \chiL^{\phantom\dag} 
\,.
\label{e.RenL}
\end{equation}
Here, the new fermions have only gauge interactions. 
In particular, they have no direct couplings to the Higgs or SM fermions, 
and $\chiL$ is completely decoupled.
The $\gu{1}_\X$ symmetry is crucial here; 
without it, we would have a dangerous renormalizable interaction \EQ{H^{\dag\!} \sigma^{a\!} \sigma^2 \ell_\LL^\dag \omega^a_\RR} 
with the SM Higgs field~$H$ and lepton doublet~$\ell_\LL$,
which would mix the new fermions and SM leptons.
To avoid severe constraints from lepton flavor violation,  
we have chosen $\omega$ to be Dirac rather than Majorana, with the conserved charge, $\gu{1}_\X$. 

\begin{figure*}[t]
\centering
\includegraphics[width=3.2in]{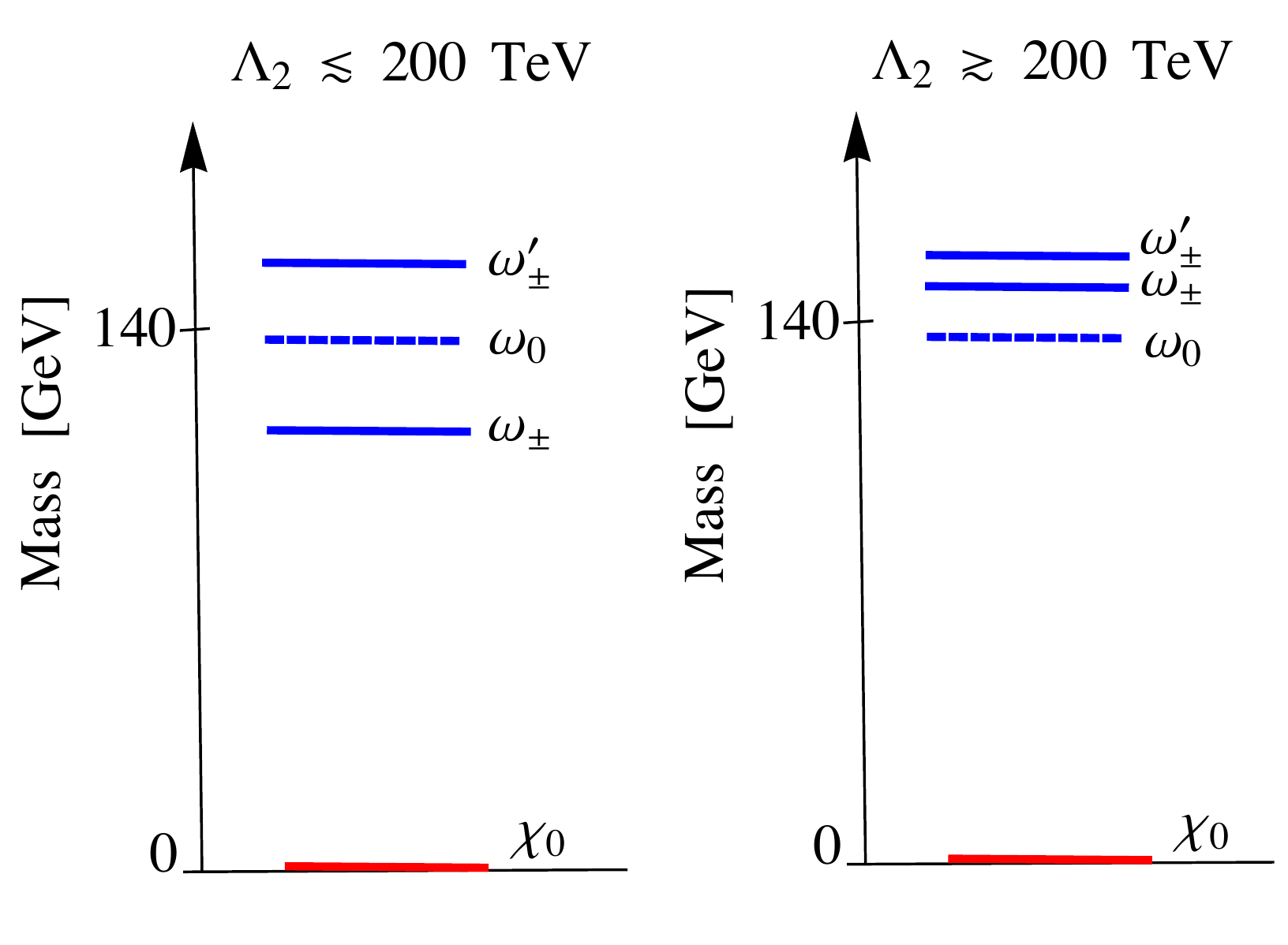}
\caption{A schematic diagram showing the mass hierarchies of the new fermions in our model with $m_\omega = 140 \gev$ for the two cases:
(i) \EQ{\Lambda_2 \simlt 200 \tev}, (ii) \EQ{\Lambda_2 \simgt 200 \tev}. 
The exact mass-splittings primarily depend on the scale $\Lambda_2$ and to a much lesser degree on the scale $\Lambda_3$ as long 
as $\Lambda_3 \simgt 10\tev$.}
\label{f.masses}
\end{figure*} 

The absence of direct couplings between the new fermions and the Higgs is an artifact of renormalizability, 
and hence vulnerable to nonrenormalizable interactions. 
At dimension-5, three non-gauge interactions arise between the new fermions and the Higgs field:
\begin{align}
\co_5^{(1)} = H^{\dag\!} H \, \omega_\RR^{a\dag} \omega_\LL^a  
\,,\quad
\co_5^{(2)} = \I \epsilon^{abc} H^{\dag\!} \sigma^{a\!} H \, \omega_\RR^{b\dag} \omega_\LL^c  
\,,\quad
\co_5^{(3)} = H^{\dag\!} \sigma^{a\!} H \, \omega_\RR^{a\dag} \chiL  
\,,
\label{e.5DOp}
\end{align}
where $\sigma^a$ are the Pauli matrices. 
Again, the $\gu{1}_\X$ symmetry plays a crucial role here 
in forbidding dangerous operators like \EQ{H^{\T\!} \sigma^{2\!} \sigma^{a\!} H \, e_\RR^\dag \omega_\LL^a}. 
The absence of direct coupling between the new and SM fermions therefore persists at dimension-5.
There are three more dimension-5 operators involving the new fermions and SM gauge field strengths:
\begin{align}
\co_5^{(4)} = g_1 B_{\mu\nu} \, \omega_\RR^{a\dag} \sigma^\mu \overline{\sigma}^\nu \omega_\LL^a    
\,,\quad
\co_5^{(5)} = \I g_2 \, \epsilon^{abc} \, W_{\mu\nu}^a \, \omega_\RR^{b\dag} \sigma^\mu \overline{\sigma}^\nu \omega_\LL^c    
\,,\quad
\co_5^{(6)} = g_2 W_{\mu\nu}^a \, \omega_\RR^{a\dag} \sigma^\mu \overline{\sigma}^\nu \chiL  
\,.
\label{e.dipoles}
\end{align}
The full Lagrangian we consider is given by 
\begin{align}
\cl = \cl_\text{SM} + \cl_\text{ren} + \sum_n \frac{\co_5^{(n)}}{\Lambda_n}
\label{e.fullL}
\end{align}
with \EQ{\Lambda_n \simgt 1 \tev}\@.

We focus on the effects of $\co_5^{(2)}$ and $\co_5^{(3)}$ for all of our studies of the LHC phenomenology 
unless otherwise noted.
$\co_5^{(1)}$ just gives a universal mass shift of order 
\EQ{v^2 / \Lambda_1} to $m_\omega$ upon EWSB, 
and induces tiny couplings, suppressed by $\Lambda_1$, between $\omega$ and the Higgs boson $h$, 
the signals from which would be swamped at the LHC by those from the renormalizable gauge interactions
unless $\Lambda_1$ is as low as the TeV scale.
We will therefore ignore $\co_5^{(1)}$ hereafter unless otherwise noted. 
In contrast, the operators $\co_5^{(2)}$ and $\co_5^{(3)}$ upon EWSB
induce mass splittings between different components of the triplet $\omega$, 
thereby opening up phase space for transitions between those states via a virtual $W^\pm$. 
Specifically, the operator $\co_5^{(2)}$ after EWSB (\EQ{v \simeq 246\gev}) gives mass terms 
\begin{align}
-\frac{v^2}{2\Lambda_2} (\omega_{-\RR}^{\prime\,\dag} \omega'_{-\LL} - \omega_{+\RR}^\dag \omega_{+\LL}^{\phantom\dag} ) + \hc
\,,
\end{align}
thereby splitting the masses of the two charged states $\omega_\pm$ and $\omega'_\mp$ by \EQ{\sim v^2 / \Lambda_2}, 
where the splitting can be as large as several GeV for \EQ{\Lambda_2 > 10\tev}. 
The neutral component $\omega_0$ receives no contributions from this operator. 
Like those from $\co_5^{(1)}$, 
the Yukawa couplings from $\co_5^{(2)}$ between $\omega$ and $h$ would be completely swamped by the renormalizable gauge interactions, 
and hence not interesting.
On the other hand, the operator $\co_5^{(3)}$ upon EWSB induces a mass term 
\begin{align}
-\frac{v^2}{2\Lambda_3} \omega_{0\RR}^\dag \chiL^{\phantom\dag} + \hc
\,,
\end{align}
thereby slightly mixing the neutral states $\omega_{0\LL}$ and $\chiL$, 
while leaving the charged components unchanged. 
The mixing induces a Yukawa coupling of the size \EQ{v / \Lambda_3} among $\omega_{0\RR}$, $\chi_0$ and $h$, 
thereby dramatically altering the phenomenology of $\chi_0$, as it was completely inert at the renormalizable level. 
The increase of the $\omega_0$ mass due to this mixing is of order \EQ{v^4 / \Lambda_3^2 m_\omega}, 
which can be at most several tens of MeV for \EQ{\Lambda_3 > 10\tev} and \EQ{m_\omega \sim 100\gev}, 
and hence its phenomenological effects are quite minimal. 
Therefore, we expect a rich collider phenomenology in a two dimensional parameter space spanned by $\Lambda_2$ and $\Lambda_3$, 
where the former controls the charged sector by creating a mass gap between $\omega_\pm$ and $\omega'_\mp$, 
while the latter governs the neutral sector by giving a $\omega_0$-$\chi_0$-$h$ Yukawa coupling.
In particular, the smallness of these mass splitting and Yukawa coupling 
gives rise to the interesting possibility that 
the transitions between the new fermion states involving $h$ and/or $W^\pm$ can occur 
with possibly \emph{macroscopic} decay lengths. 

We will ignore the dipole operators~$\co_5^{(4,5,6)}$, with the following justifications. 
First, the following very simple UV completion of the effective Lagrangian~\erefn{fullL} 
suggests that it is natural to take \EQ{\Lambda_{4,5,6} \gg \Lambda_{1,2,3}}.
Imagine a heavy Dirac fermions $\psi$ with SM gauge quantum numbers \EQ{(\mathbf{1}, \mathbf{2})_{1/2}} 
and a $\gu{1}_\X$ charge $+1$. 
The most general renormalizable lagrangian involving $\psi$ is given by 
\begin{align} 
\I \overline{\psi} \gamma^\mu \D{\mu} \psi
-M \overline{\psi} \psi
-(\lambda \, H^{\dag\!} \sigma^a \omega_\RR^{a\dag} \psi_\LL^{\phantom\dag} 
+\lambda' \, \psi_\RR^\dag \omega^a_\LL \sigma^{a\!} H 
+\lambda'' \, \psi_\RR^\dag \chiL H + \hc)
\,,
\label{e.UVcompletion}
\end{align}
where the $\gsu{2}_\LL$-doublet indices among $\sigma^a$, $\psi$ and $H$ are implicit.
Upon integrating out the heavy fermion $\psi$, 
$\co_5^{(1,2,3)}$ are generated at tree level 
with \EQ{\Lambda_1 = \Lambda_2 = M / \lambda \lambda'} and \EQ{\Lambda_3 = M / \lambda \lambda''}.
On the other hand, 
the dipole operators~$\co_5^{(4,5,6)}$ are generated at 1-loop with  
\EQ{\Lambda_{4,5,6} \sim 16\pi^2 \Lambda_{1,2,3}}, respectively. 
Thus, $\co_5^{4,5,6}$ are suppressed by two more orders of magnitude than $\co_5^{1,2,3}$ in the effective Lagrangian~\erefn{fullL}. 
The effects of $\co_5^{(4,5)}$ are then completely swamped by those of the renormalizable gauge interactions, so we will ignore 
$\co_5^{(4,5)}$ hereafter.
The operator $\co_5^{(6)}$ could induce a potentially interesting process $\omega_0 \to \chi_0 + \gamma/Z$,%
\footnote{Prompt and displaced $\gamma/Z$ signatures also arise in GMSB model ~\cite{GMSB, GMSB:Displaced, NeutralinoOscillation}.}
although with the $(16\pi^2)^2$ suppression, it would be completely subdominant relative to the $\co_5^{(3)}$-induced process, 
$\omega_0 \to \chi_0 + h$, unless we (almost) close the phase space for the latter by having $m_\omega$ nearly at or below $m_h$. 
Since we are interested in robust on-shell Higgs production in this paper and will take $m_\omega$ sufficiently above $m_h$, 
we will also ignore $\co_5^{(6)}$ hereafter. 

We are concerned with small mass splittings;
therefore it is important to take into account mass splitting effects induced by $W^\pm$ and $Z$ loops, 
which arise even at the renormalizable level.
In the absence of nonrenormalizable operators, 
$\omega_\pm$, $\omega'_\mp$ and $\omega_0$ are degenerate at tree level by the $\gsu{2}_\LL$ symmetry.
However, since electroweak symmetry is broken, we expect that the degeneracy should be lifted by loops.
The one-loop radiative mass splittings between $\omega_0$, $\omega_\pm$ and $\omega'_\mp$ 
from the renormalizable gauge interactions alone 
are given by~\cite{MassSplittingWino, MDM}:
\begin{align}
\Delta m_{\omega} 
&\equiv 
m_{\omega_\pm} - m_{\omega_0} 
= 
m_{\omega'_\mp} - m_{\omega_0}
\nn\\
&=
\frac{\alpha_2}{4 \pi} \, m_\omega 
\!\left[ 
f \!\left( \frac{m_\W}{m_\omega} \right) 
-  
f \!\left( \frac{m_\Z}{m_\omega} \right) 
\cos^2\!\thetaW
\right],
\label{e.mass-splitting}
\end{align}
where
\begin{equation}
f(r) \equiv r^4 \log{r} + r(2+r^2) \sqrt{4 - r^2} \, \mathrm{Arctan} \!\left( \frac{\sqrt{4-r^2}}{r} \right)
\end{equation}
Here, \EQ{\Delta m_\omega} is always positive, rendering $\omega_0$ lighter than $\omega_\pm$ and $\omega'_\mp$ 
(in the absence of the contributions from $\co_5^{(2)}$).  
Recently, \EQ{\Delta m_\omega} has been calculated at two-loop level~\cite{2loopWino}, 
which significantly reduces the renormalization scale dependence compared to one-loop computation. 
Numerically, the mass splitting varies between $\sim 150$--$165\mev$ for $m_\omega$ between $\sim 100$--$1000\gev$. 

To summarize, the phenomenological Lagrangian of our model is
\begin{equation}
\cl = \cl_\text{SM} + \cl_\text{kin} + \cl_\text{mass} + \cl_\text{gauge} + \cl_\Lambda
\,,\label{e.Lpheno} 
\end{equation}
where, at the leading non-trivial orders in \EQ{v / \Lambda_{2,3}}, 
\begin{align}
\cl_\text{kin} 
&=
\I \overline{\omega}_+ \gamma^{\mu} \partial_\mu \omega_+ 
+ \I \overline{\omega}'_- \gamma^{\mu} \partial_\mu \omega'_- 
+ \I \overline{\omega}_0 \gamma^{\mu} \partial_\mu \omega_0 
+ \I \chiL^\dag \overline{\sigma}^{\mu} \partial_\mu \chiL^{\phantom\dag} 
\,,\\
\cl_\text{mass} 
&=
- m_{\omega_\pm\,} \overline{\omega}_+ \omega_+
- m_{\omega'_\mp\,} \overline{\omega}'_- \omega'_-
- m_{\omega_0\,} \overline{\omega}_0 \omega_0
\,,\\
\cl_\text{gauge}
&=
g_2 \!\left[ 
W_{\mu}^+ 
\!\left( 
\overline{\omega}_+ \gamma^\mu \omega_0 
-\overline{\omega}_0 \gamma^\mu \omega'_- 
\right) 
+\hc \right] 
-g_2 \cos\thetaW \, Z_\mu 
\!\left( 
\overline{\omega}_+ \gamma^\mu \omega_+ 
-\overline{\omega}'_- \gamma^\mu \omega'_- 
\right)
\label{e.Lgauge}
\nn\\%
&\phantom{=}
-e A_\mu \!\left( 
\overline{\omega}_+ \gamma^\mu \omega_+ 
-\overline{\omega}'_- \gamma^\mu \omega'_- 
\right) 
,\\
\cl_\Lambda
&=
-g_2 \frac{v^2}{2 \Lambda_3 m_\omega} 
\!\left[ W_\mu^+ 
\bigl( 
\omega_{+\LL}^\dag \overline{\sigma}^\mu \chiL^{\phantom\dag} 
-\chiL^\dag \overline{\sigma}^\mu \omega'_{-\LL} 
\bigr)
+\hc \right] 
-\frac{v}{\Lambda_3} h 
\bigl(
\omega_{0\RR}^\dag \chiL^{\phantom\dag} 
+\chiL^\dag \omega_{0\RR}^{\phantom\dag}
\bigr)
\,,\label{e.L_Lambda}
\end{align}
with
\begin{align}
m_{\omega_\pm} = m_\omega + \Delta m_\omega - \frac{v^2}{2\Lambda_2}
\,,\quad
m_{\omega'_\mp} = m_\omega + \Delta m_\omega + \frac{v^2}{2\Lambda_2}
\,,\quad
m_{\omega_0} = m_\omega + \frac{v^4}{8 m_\omega \Lambda_3^2}  
\,.
\label{e.masses}
\end{align}
Here, we have chosen $\Lambda_3$ to be real and positive without loss of generality by adjusting the phase of $\chiL$. 
The phase of $\Lambda_2$ cannot be removed by field redefinition, however, and it would 
lead to CP-violating Yukawa couplings between $\omega$ and $h$. 
But since it would be hard to observe such CP violation effects, 
we have taken $\Lambda_2$ to be also real and positive in~\eref{masses} for simplicity.

Before we move on to the discussion of collider phenomenology of the Lagrangian~\erefn{Lpheno}, 
we would like to make a few comments on possible variations of the model. One simple variation is  
to get rid of $\gu{1}_\X$ and make $\omega$ Majorana instead of Dirac, though this possibility has  flavor issues as discussed earlier. 
Another possibility is to keep $\gu{1}_\X$ and make $\chi_0^\PD$ also Dirac by adding $\chi_{0\RR}^\PD$ to the theory. 
These two variations have difficulties in cosmology, however. 
In the Majorana case, fine-tuning would be necessary to make $\chi_0$ massless, 
while in the Dirac $\chi_0$ case, there is no reason for $\chi_0$ to be massless.
A massive $\chi_0$ would lead to over-closure of the universe unless its mass is \EQ{\simlt 1\>}eV, 
since $\chi_0$ is extremely weakly interacting and stable. In fact, the mass must be \EQ{\ll 1\>}eV, 
because if it is near \EQ{\sim 1\>}eV, 
(a significant fraction of) dark matter would be hot, thus at odds with observation.
Therefore, the simplest approach is to prevent $\chi_0$ from acquiring mass by symmetry as in our model.

The big-bang nucleosynthesis (BBN) bound on extra relativistic degrees of freedom do not necessarily constrain the 
massless $\chi_0$, because
the $\chi_0$ gas is not in equilibrium with the SM gas unless the temperature is far above the weak scale.
Thus, the temperature of $\chi_0$ can be much lower than that of the SM during BBN\@.
Also, between the weak scale and the BBN temperature, the SM gas gets heated up several times, 
possibly through electroweak phase transition, and certainly via the annihilations of massive particles such as $t$, $Z/W^\pm$, $b$, etc., 
as the temperature drops below their masses. The $\chi_0$ gas, on the other hand, remains unheated.

\section{Collider Phenomenology}
\label{s.Pheno}
In this section, we discuss the collider phenomenology of the lagrangian~\erefn{Lpheno}. 
We first study the production mechanisms of the new fermions in our model and their decay channels and widths. 
In particular, we emphasize that long-lived particles decaying to Higgs plus missing energy is a prominent signature of our model 
in a large portion of the parameter space. 
We will then divide our parameter space into eight regions depending on the signatures, and 
discuss for each region the current experimental constraints on our model as well as discovery prospects. 
We will see that some of the exotic events would have already been selected by the ATLAS and CMS triggers 
in their $7$ and $8\tev$ runs and can be used for possible discovery. 

\subsection{$\omega$ production, decay, and the benchmark point}
The new fermions $\omega$ can be Drell-Yan produced at hadron colliders through gauge interaction in 
$\cl_\text{gauge}$~\eref{Lgauge} as: 
\begin{align}
p\,p \to \gamma/Z^{*} \to \omega_{+} \, \omega_{-}
&\,,\quad 
p\,p \to \gamma/Z^{*} \to\omega_{-}' \, \omega_{+}'
\,,\nn\\
p\,p \to W^{+*} \to \omega_{+} \, \overline{\omega}_{0} 
&\,,\quad
p\,p \to W^{+*} \to \omega_{0} \, \omega_{+}'
\,,\nn\\ 
p\,p \to W^{-*} \to \omega_{0} \, \omega_{-}
&\,,\quad
p\,p \to W^{-*} \to \omega'_{-} \, \overline{\omega}_{0} 
\,.\nn
\end{align}
(At $e^+ e^-$ colliders, only the $\gamma/Z^*$ mediated processes are possible.)
Recall the $\gu{1}_\X$ symmetry under which $\omega_+$, $\omega'_-$ and $\omega_0$ carry charge $+1$, while 
their antiparticles $\omega_-$, $\omega'_+$ and $\overline{\omega}_0$ carry $-1$, thus forbidding  
the final states like \EQ{\omega_0 \, \omega_+} as the SM particles are all $\gu{1}_\X$ neutral.
Although, direct $\chi_0$ production at the colliders is highly suppressed by \EQ{v / \Lambda_3}, 
the fermions $\omega$ eventually decay to $\chi_0$, which is the lightest stable particle (LSP) in our model and 
appears as missing energy in the detector.

Clearly, the signatures of our model depend on the decay of $\omega$ fermions. 
In~\eref{masses}, we observe that the effect of the scale $\Lambda_3$ on the mass of the neutral particle $\omega_0$ is at most a few 
tens of MeV for $\Lambda_3 \gg 10\tev$ and $m_\omega \sim 100\gev$. 
Therefore, the mass hierarchy is more or less determined by the scale $\Lambda_2$, with $\omega'_\mp$ being the heaviest 
among the three $\omega$ fermions%
\footnote{For $\Lambda_3 \simlt 10\tev$, it is possible for the charged fermions, $\omega_\pm$ and $\omega_\pm'$ to be lighter than 
the neutral $\omega_0$. However, in this case, all the $\omega$ fermions decay promptly to $\chi_0$, since $v / \Lambda_3$ is not small.  
This signature is phenomenologically no different from the \emph{Region~(iv)} signals to be discussed in section~\ref{s.EightRegions}.} 
while the next-to-lightest stable particle (NLSP) could be either $\omega_0$ or $\omega_\pm$. 
A schematic diagram for the mass hierarchies of the $\omega$ fermions is shown in \fref{masses} for $m_\omega = 140 \gev$, 
which will be our \emph{benchmark point} for the rest of the discussion unless otherwise stated.%
\footnote{A reason for this benchmark point is to have \EQ{m_{\omega_0} > m_h} such that the Higgs boson $h$ 
in \EQ{\omega_0 \to \chi_0 \, h} is always produced on-shell.}
Led by this observation of mass hierarchy among the $\omega$ fermions, we can roughly divide the parameter space of interest 
into two categories:
\begin{itemize}
\item[(I)]{
\underline{$\omega_\pm$ NLSP}:\\ 
For $\Lambda_2 \simlt 200 \tev$, the mass formulas~\erefn{masses} tell us that the \EQ{v / \Lambda_2} effects, which tries to 
lower $m_{\omega_\pm}$ below $m_{\omega_0}$, is greater than the radiative mass splitting \EQ{\Delta m_\omega}, 
which tries to push $m_{\omega_\pm}$ above $m_{\omega_0}$.
Consequently, $\omega_\pm$ is the NLSP with mass-splitting \EQ{m_{\omega_0} - m_{\omega_\pm}} 
as big as a few GeV for $\Lambda_2 \simlt 20 \tev$. 
The only decay channel for the NLSP is \EQ{\omega_+ \to \chi_0 \, W^+} (or the charge-conjugate process 
\EQ{\omega_- \to \overline{\chi}_0 \, W^-}) 
with decay width
\begin{equation}
\Gamma (\omega_+ \to \chi_0 \, W^+) 
= 
m_{\omega_\pm} \frac{v^2}{32 \pi \Lambda_3^2} 
\!\left( 1 - \frac{m_\W^2}{m_{\omega_\pm}^2} \right)^{\!\!2} 
\!\left( 1 + 2 \frac{m_\W^2}{m_{\omega_\pm}^2} 
\right).
\label{e.decay:chi0+W}
\end{equation}
On the other hand, the neutral $\omega$ has two decay modes available:
\begin{itemize}
\item[(i)]{\EQ{\omega_0 \to \chi_0 \, h} (or the charge-conjugate process \EQ{\overline{\omega}_0 \to \overline{\chi}_0 \, h}) 
with a decay width
\begin{equation}
\Gamma ( \omega_0 \to \chi_0 \, h) 
=  
m_{\omega_0} \frac{v^2}{32 \pi  \Lambda_3^2} 
\!\left( 1 - \frac{m_h^2}{m_{\omega_0}^2} \right)^{\!\!2},
\label{e.decay:higgs}
\end{equation}
}
\item[(ii)]{\EQ{\omega_0 \to \omega_+ \, W^{-*}} (or the charge-conjugate process \EQ{\overline{\omega}_0 \to \omega_- \, W^{+*}}) 
with
\begin{equation}
\begin{split}
&
\Gamma (\omega_0 \to \omega_+ \, \ell^- \, \bar{\nu}_\ell ) 
= \frac{2}{15 \pi^3} G_\text{F}^2 \, \delta m_\omega^5 \, 
F\!\left( \frac{m_\ell^2}{\delta m_\omega^2} \right) 
\Theta(\delta m_\omega - m_\ell)
\,,\\
& 
\Gamma (\omega^0 \to \omega_+ + \text{hadrons})
\\%
&\quad 
= 
\begin{cases}
0 
&; \delta m_\omega \leq m_{\pi^+}
\\%
\displaystyle{ 
\Gamma_{ \omega^0 \to \omega_+ \, \pi^- } 
=
\frac{4 G_\text{F}^2 |V_{ud}|^2 f_{\pi}^2}{\pi} \,
\delta m_\omega^3 
\!\left( 1 - \frac{ m_{\pi^\pm}^2}{\delta m_\omega^2} \right)^{\!\!1/2}
} 
&; m_{\pi^+} < \delta m_\omega \simlt m_\rho
\\%
\displaystyle{
\mathrm{max} 
\!\left\{ 
\Gamma_{ \omega^0 \to \, \omega_+ \, \pi^- }
,\> 
\kappa N_\C \Gamma_{ \omega_0 \to \, \omega_+ \, e^- \, \bar{\nu}_e }
\right\}
} 
&; \delta m_\omega \simgt m_\rho
\end{cases}
\label{e.weakdecays}
\end{split}
\end{equation}
where \EQ{\delta m_\omega \equiv m_{\omega_0} - m_{\omega_\pm}}, \EQ{N_\C = 3}, 
$f_{\pi} \simeq 92.2 \mev$ is the pion decay constant, 
$G_\text{F}$ is the Fermi constant, and the function $F$ is given by
\begin{equation}
F(a) 
\equiv 
\sqrt{1-a} \left( 1 - \frac{9}{2} a - 4a^2 \right)
+ \frac{15}{2} a^2 \, \mathrm{Arctanh} \sqrt{1-a}
\end{equation}
The coefficient $\kappa$ appearing in \eref{weakdecays} is an $\co(1)$ factor 
introduced to parametrize the uncertainty in the hadronic decay width of $\omega$, which we estimate to be 
on the order of the color factor $N_\C$ times the leptonic decay rate, 
\EQ{\Gamma_{ \omega_0 \to \, \omega_+ \, e^- \, \bar{\nu}_e }}. 
In our analyses below, $\kappa$ will be varied between $0$ and $2$. 
}
\end{itemize}
The heaviest of the new fermions $\omega'_\mp$ can decay to lighter states $\omega_0$, $\overline{\omega}_0$, $\chi_0$ and 
$\overline{\chi}_0$ through the channels, 
\EQ{\omega'_- \to \chi_0 \, W^-}, 
\EQ{\omega_-' \to \omega_0 \, W^{-*}} and their charge-conjugate processes. 
Their decay widths can be calculated by the replacements $m_{\omega_\pm} \to m_{\omega'_\mp}$ in \eref{decay:chi0+W}, and 
$\omega_0 \to \omega_-'$ and $\omega_+ \to \omega_0$ in \eref{weakdecays}. 
Note that the \emph{direct} decay of $\omega_\pm'$ to $\omega_\pm$ is forbidden by $\gu{1}_\X$ symmetry 
(but it can happen in two-step weak decays with an intermediate $\omega_0$).
} 
\item[(II)]{
\underline{$\omega_0$ NLSP}:\\ 
If the scale $\Lambda_2 \simgt 200 \tev$, 
the mass splittings between the $\omega$ fermions are dominated by \EQ{\Delta m_\omega}.
In this scenario, the NLSP is $\omega_0$, 
while $\omega_\pm$ and $\omega'_\mp$ are nearly degenerate with each other for \EQ{\Lambda_2 \gg 10^3 \tev}. 
The only decay channel of the NLSP is \EQ{\omega_0 \to \chi_0 \, h}, 
while the fermions $\omega_\pm$ and $\omega'_\mp$ have additional decay modes 
\EQ{\omega_+ \to \omega_0 \, W^+} and \EQ{\omega'_- \to \omega_0 \, W^-} besides 
\EQ{\omega_+ \to \chi_0 \, W^+} and \EQ{\omega'_- \to \chi_0 \, W^-}, respectively. 
The decay widths for $\omega_0$, $\omega_\pm$ and $\omega'_\mp$ can be deduced from 
\erefs{decay:chi0+W}{weakdecays} by appropriate replacements as in the previous case.
Since $\omega'_\mp$ is nearly degenerate with $\omega_\pm$, 
its decay widths are the same as the corresponding widths of $\omega_\pm$.
} 
\end{itemize}
\begin{figure*}[t]
\centering
\begin{tabular}{ccc}
(a) $\omega_0$ & (b) $\omega_\pm$ & (c) $\omega_\mp'$\\
\includegraphics[width=2in]{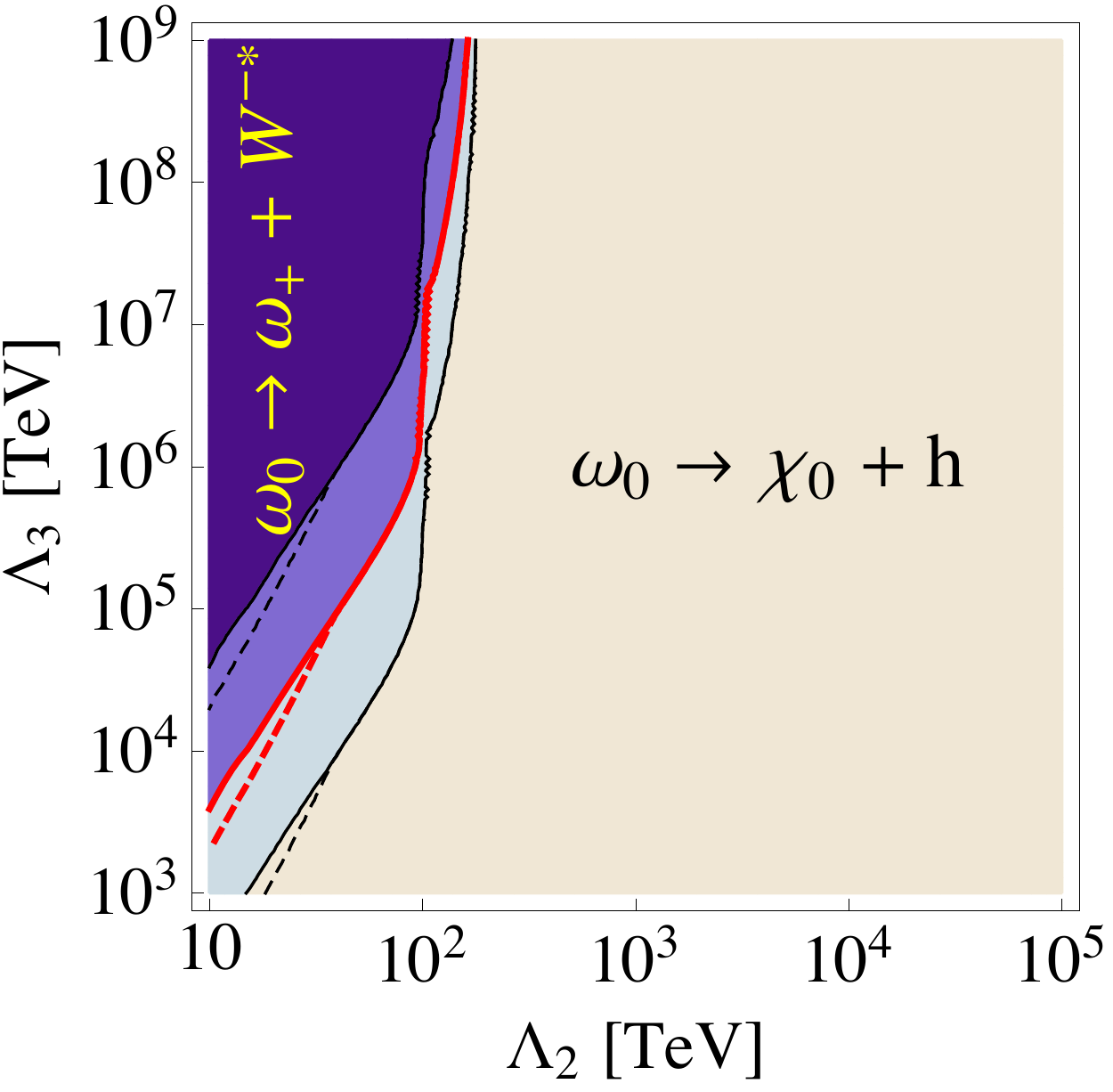}
&\includegraphics[width=2in]{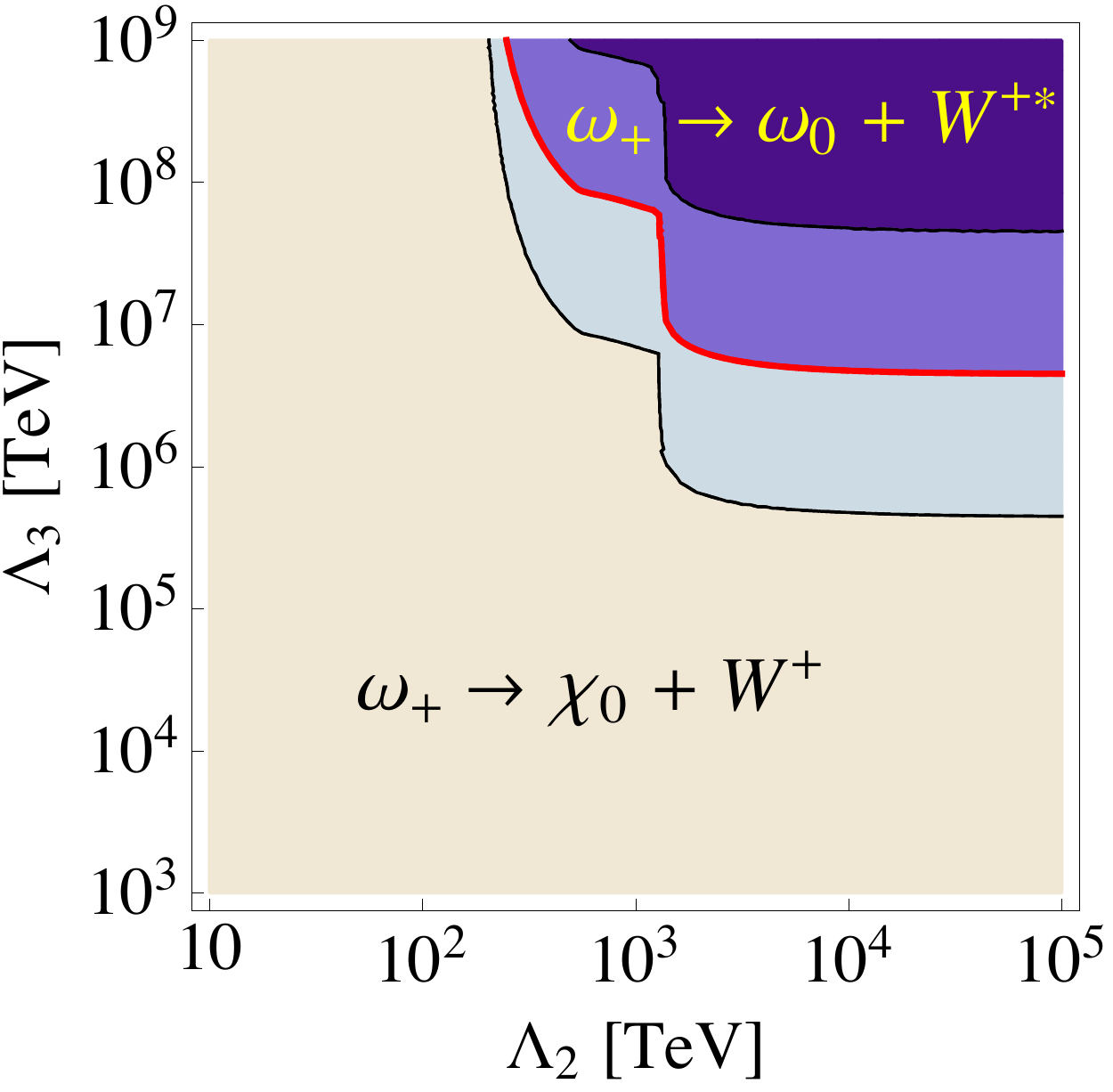}
&\includegraphics[width=2in]{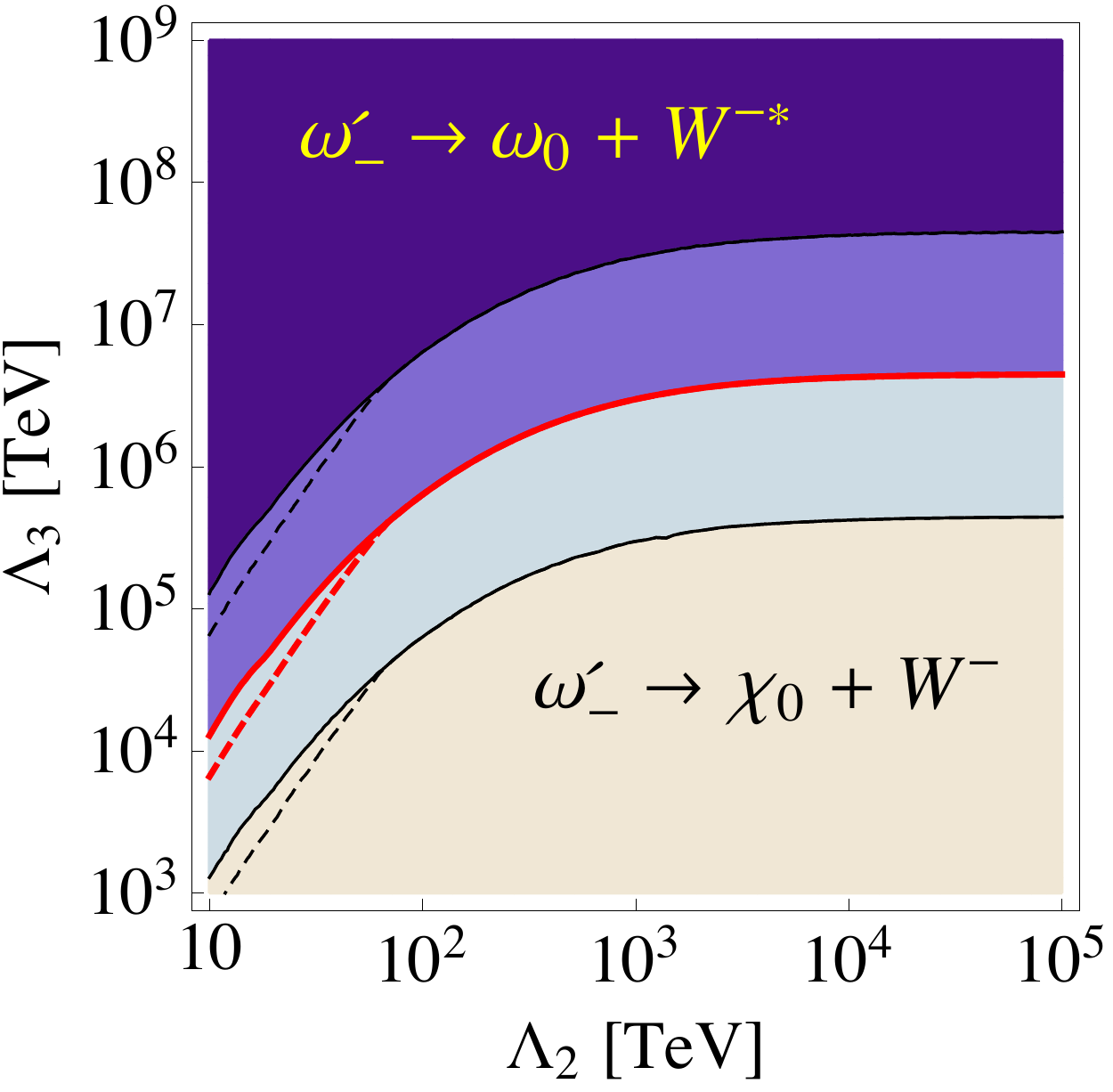}
\\
\includegraphics[width=2in]{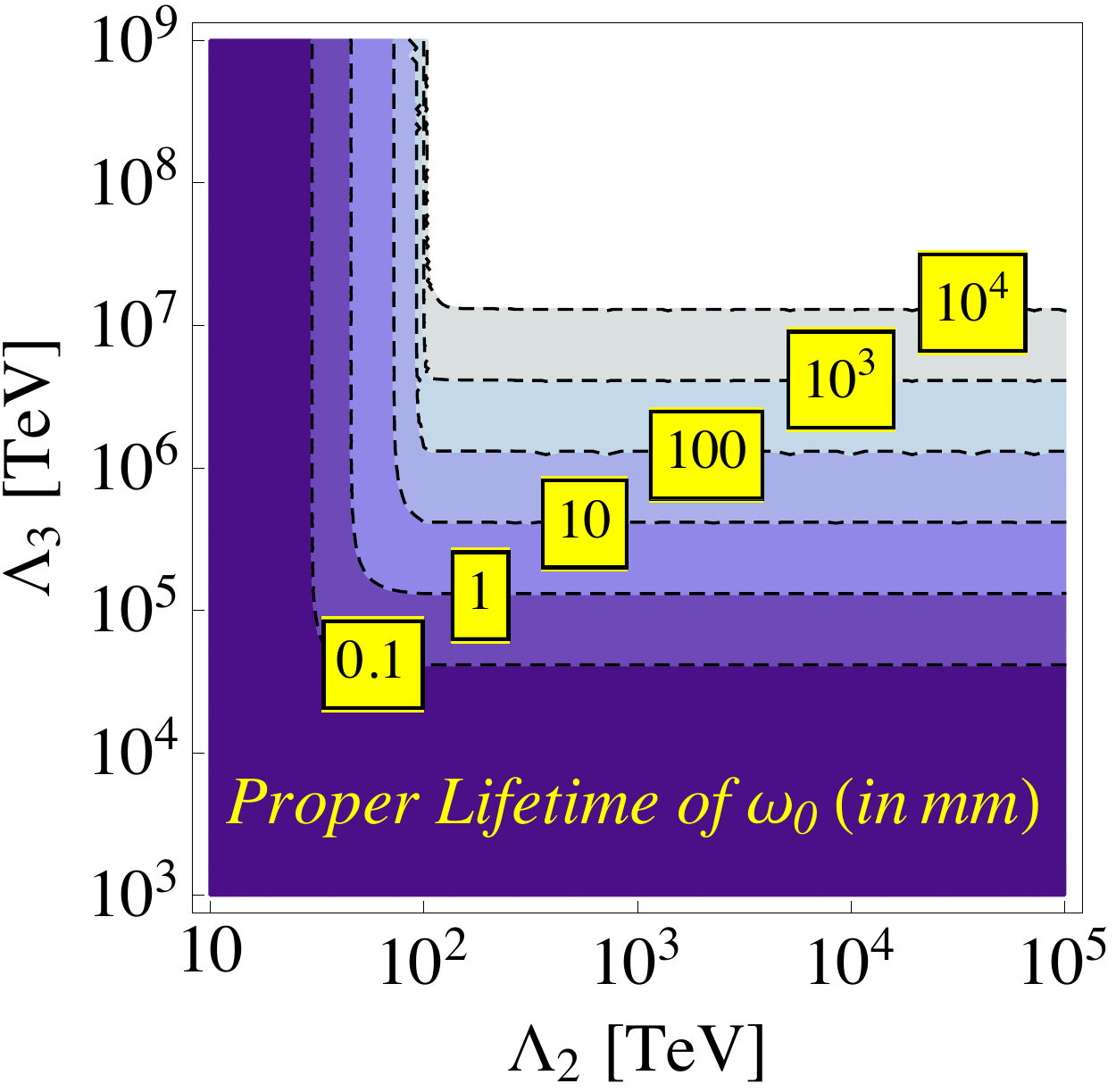}
&\includegraphics[width=2in]{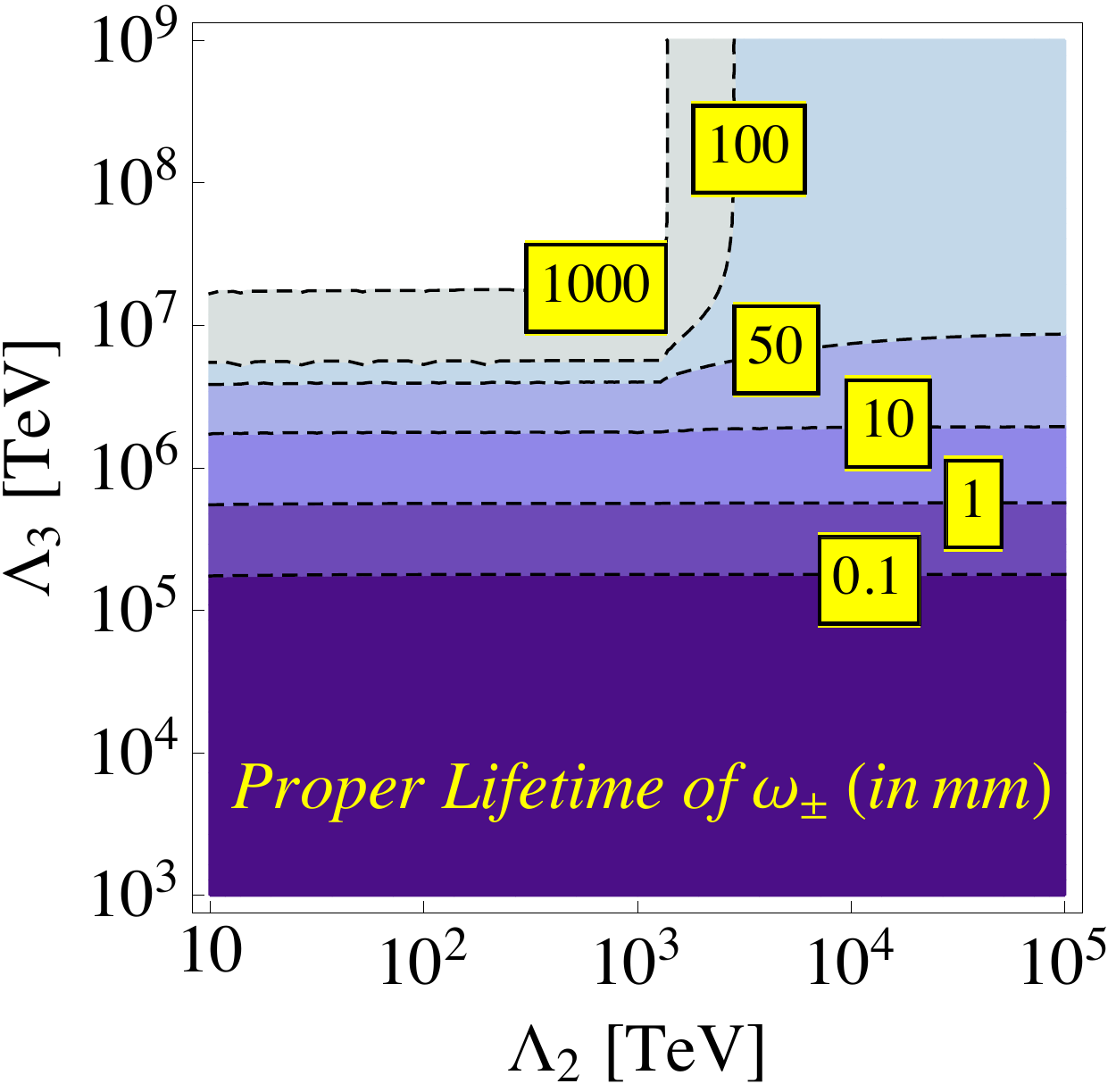}
&\includegraphics[width=2in]{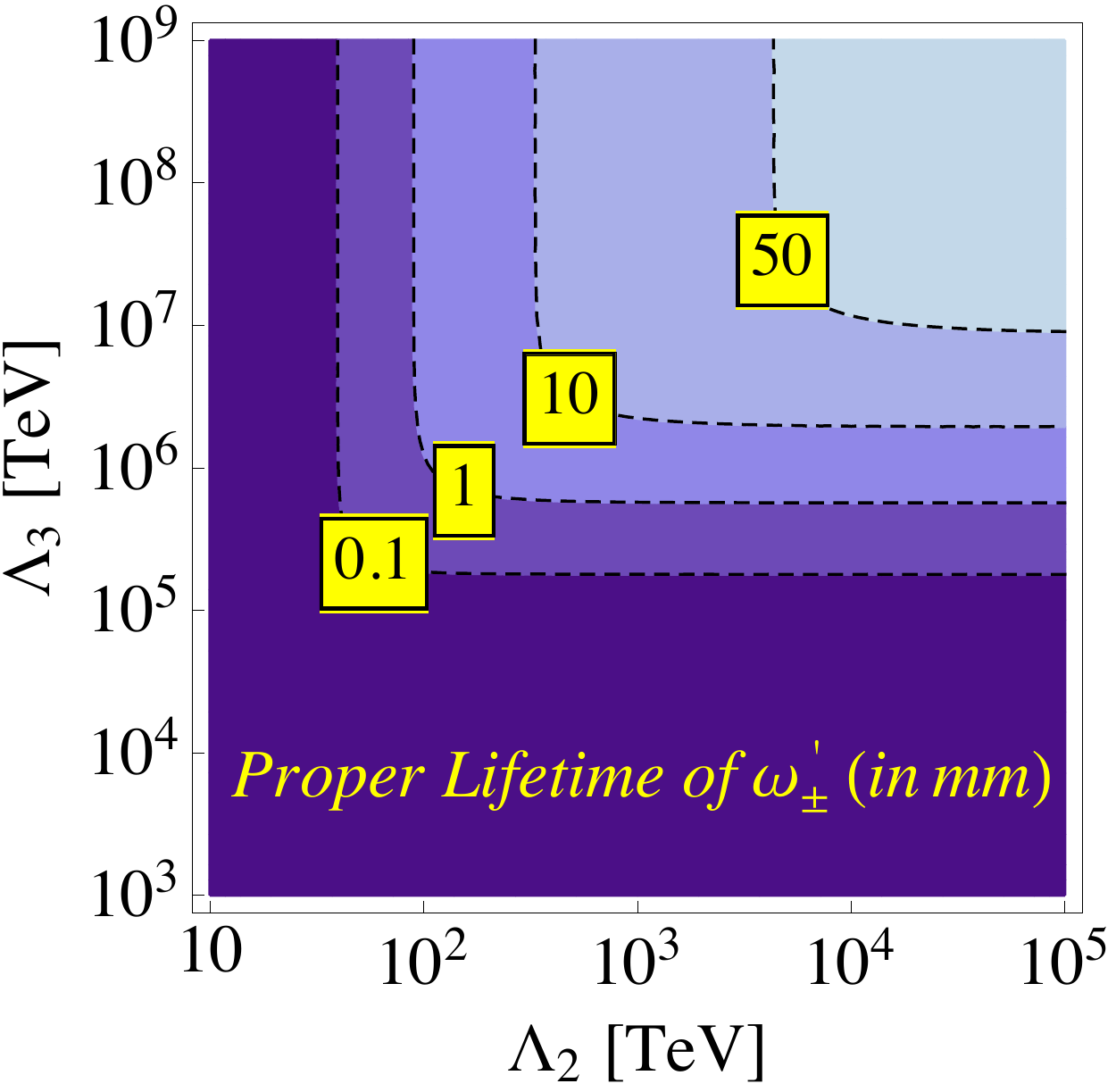}
\end{tabular} 
\caption{
\emph{\underline{Top Row}}: 
Branching Ratio ($BR$) contour curves are plotted for (a) $\omega_0$, (b) $\omega_\pm$ and (c) $\omega_\mp'$ fermions 
in the $\Lambda_2$-$\Lambda_3$ parameter space for the benchmark point \EQ{m_\omega = 140 \gev}. 
The central solid red line depicts a $BR=50\,\%$ for the two decay channels. 
The two solid black lines correspond to $BR=99\,\%$ for each of the two decay channels. 
Finally, the dashed lines for each $BR$ contour line depicts the uncertainty in the calculation of hadronic decays 
by varying the $\kappa$ parameter between $0$--$2$ (see text for definition of $\kappa$). 
\emph{\underline{Bottom Row}}: 
Lifetimes of the $\omega$ fermions as contours in the $\Lambda_2$-$\Lambda_3$ parameter space for $m_\omega = 140 \gev$.}
\label{f.scans}
\end{figure*} 
In \fref{scans}, the branching ratio and decay lifetimes of the $\omega$ fermions are plotted in the $\Lambda_2$-$\Lambda_3$ parameter space for our benchmark point. A few comments are in order:
\begin{itemize}
\item{
The lifetime of the heavier charged fermion $\omega_\mp'$, shown in~\fref{scans}(c), never exceeds $\approx 100$~mm. 
Naively, one might expect the lifetime to increase as the scale $\Lambda_3$ increases and suppresses the 
$W^\pm$-$\chi_0$-$\omega_\mp'$ coupling~\eref{L_Lambda}. 
However, since the splitting between $\omega_0$ and $\omega_\mp'$ remains \EQ{\simgt 150 \mev} 
due to the radiative contribution~\eref{mass-splitting}, 
there is always a phase space for the weak decay \EQ{\omega'_- \to \omega_0 \, W^{-*}}, keeping the $\omega'_\mp$ lifetime below
$\approx 100$~mm. The leptons/hadrons from this $W^{-*}$ will be too soft to be detected at the LHC or Tevatron. Consequently, 
the $\omega'_\mp$ production will just appear as the $\omega_0$ production for all practical purposes.
} 
\item{ 
The same, however, does not hold true for the lighter charged fermion $\omega_\pm$. Towards the upper-left corner of 
\fref{scans}(b), $\omega_\pm$ becomes the NLSP, with the only possible decay mode being \EQ{\omega_+ \to \chi_0 \, W^+}. 
The lifetime of this decay increases as $\Lambda_3$ increases and suppresses the $\omega_\pm$-$\chi_0$-$W^\mp$ coupling~\eref{L_Lambda}, 
letting the $\omega_\pm$ travel deep into the detector with lifetime $\simgt 500$~mm. 
Towards the upper-right corner of \fref{scans}(b), the NLSP becomes $\omega_0$ with the mass splitting between $\omega_0$ and $\omega_\pm$
dominated by the radiative effect~\eref{mass-splitting}, i.e., \EQ{m_{\omega_\pm} - m_{\omega_0} \approx 150 \mev}.
Here, like the $\omega'_\mp$ case above, the $\omega_\pm$ production will be equivalent to the $\omega_0$ production for all practical 
purposes.
} 
\item{
The neutral fermion, $\omega_0$ can be comparatively long-lived as shown in \fref{scans}(a). 
This follows from the fact that $\omega_0$ is the NLSP in majority of the parameter space (\EQ{\Lambda_2 \simgt 200 \tev}) 
and hence the only decay channel available is \EQ{\omega_0 \to \chi_0 \, h}. 
For $\omega_0$ lifetimes exceeding $\sim 10$~mm, its decay results in \emph{Higgs production amidst the detector} 
leading to a plethora of collider signatures as will be discussed next.
} 
\end{itemize}
%

\subsection{Simulation}
The events for $\omega$ pair production were generated in {\tt MadGraph}~\cite{MG5} and showered in {\tt Pythia}~\cite{Pythia}. 
The input model file for {\tt MadGraph} was generated using the {\tt Mathematica} package {\tt FeynRules}~\cite{FeynRules}. 
The $\omega$ pair-production cross-sections, shown in \tref{xsection}, were obtained using \tt{MadGraph}\rm. 

\begin{table}[h]
\begin{center}
\begin{tabular}{ | l | c | c | c | c |}
\hline 
 & \EQ{\sqrt{s}=1.96 \tev} & \EQ{\sqrt{s}=7 \tev} & \EQ{\sqrt{s}=8 \tev} & \EQ{\sqrt{s}=14 \tev} 
\\
 & Tevatron Run & LHC Run &  LHC Run & LHC Run
\\
\hline
\EQ{p\,p(\overline{p}) \to \omega_{0} \, \omega_{-}}             & $0.16$ pb & $0.61$ pb & $0.78$ pb & $2.01$ pb \\
\EQ{p\,p(\overline{p}) \to \overline{\omega}_{0} \, \omega_{+}}  & $0.16$ pb & $1.23$ pb & $1.52$ pb & $3.45$ pb \\
\EQ{p\,p(\overline{p}) \to \omega_{0} \, \omega_{+}'}            & $0.16$ pb & $1.23$ pb & $1.52$ pb & $3.45$ pb \\
\EQ{p\,p(\overline{p}) \to \overline{\omega}_{0} \, \omega_{-}'} & $0.16$ pb & $0.61$ pb & $0.78$ pb & $2.01$ pb \\
\EQ{p\,p(\overline{p}) \to \omega_{+} \, \omega_{-}}             & $0.22$ pb & $0.91$ pb & $1.14$ pb & $2.74$ pb \\
\EQ{p\,p(\overline{p}) \to \omega_{+}' \, \omega_{-}'}           & $0.22$ pb & $0.91$ pb & $1.14$ pb & $2.74$ pb \\
\hline
\end{tabular}
\end{center}
\caption{Cross-sections for pair-production of $\omega$ fermions at the LHC and the Tevatron for 
the benchmark point \EQ{m_{\omega} \simeq 140 \gev}.
}
\label{t.xsection}
\end{table}

While the proper lifetimes of the $\omega$ fermions has been plotted in \fref{scans}, the actual decay length inside the detector 
will, event by event, be different from the proper lifetime depending on the boost factor. 
To study this effect, we have performed an elementary detecter simulation to calculate 
the fraction of $\omega$ particles that decay in different parts of the LHC detector. 
The lifetime of a particle in the lab frame, $\tau$ is related to its proper lifetime, $\tau_0$ by
\begin{equation}
\tau = \frac{\tau_0}{\sqrt{1-v^2}} \equiv \gamma \tau_0 \,.
\end{equation}
The mean transverse distance $L_\T$ travelled by the particle in the lab frame is 
\begin{equation}
L_\T = v_\T^\PD \tau = v_\T^\PD \gamma \tau_0 = \frac{p_\T^\PD}{m}  \tau_0  \,.
\end{equation}
The pre-factor in the above equation $p_\T^\PD / m$ is precisely the distribution plotted in \fref{detector}(a). 
If $N_0$ is the number of long-lived particles produced at the primary vertex, 
the fraction $N(x_\T^\PD)$ of these particles decaying within a distance $x_\T^\PD$ in the transverse direction 
is given by
\begin{equation}
N(x_\T^\PD) = \left( 1 - \mathrm{e}^{- x_\T^\PD / L_\T} \right)\! N_0
\end{equation}
The fraction of $\omega_0$ decaying inside different sections of the ATLAS detector 
are shown in \fref{detector}(b) for \EQ{\Lambda_3 \gg \Lambda_2} (i.e., the $\omega_0$ NLSP region). 
Although our simulation has used the dimensions of ATLAS detector, the results also apply to the CMS detector to a good approximation. 
Note that $\omega_0$ with lifetimes as long as $\sim 10$~m can have significant decays inside the inner detector.  

\begin{figure*}[t]
\centering
\begin{tabular}{cc}
\includegraphics[width=2.9in]{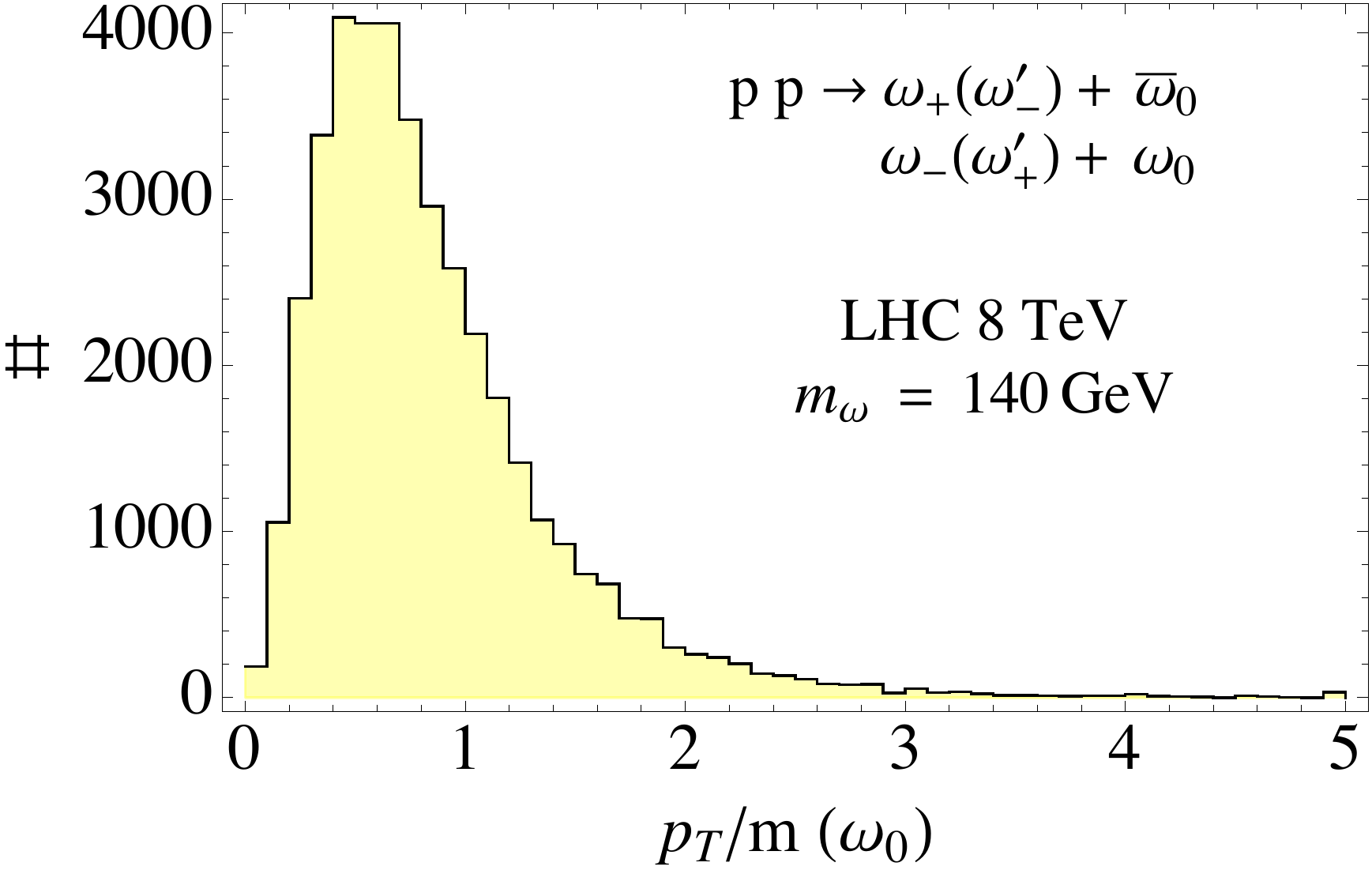}
&\includegraphics[width=3.5in]{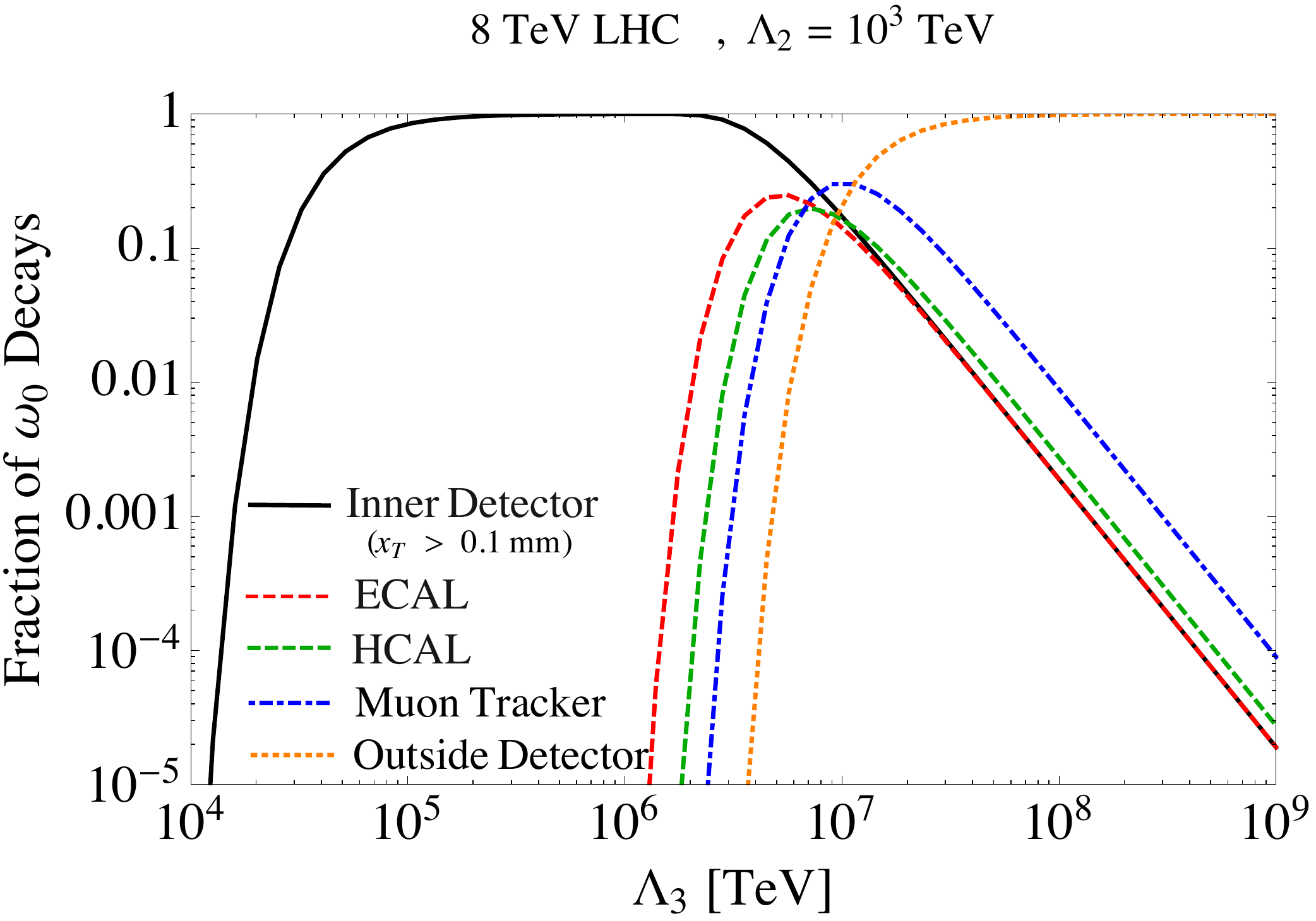}
\\
(a) & (b)
\end{tabular} 
\caption{
(a) $p_\T / m$ distribution of $\omega_0$ for $8\tev$ LHC run for \EQ{m_\omega = 140\gev} and \EQ{\Lambda_3 = 10^3 \tev} 
(i.e., the $\omega_0$ NLSP region). The normalization on the vertical axis is arbitrary. 
(b) Fraction of $\omega_0$ decaying inside the various components of the ATLAS detector as a function of \EQ{\Lambda_3} for 
\EQ{m_\omega = 140\gev} and \EQ{\Lambda_2 = 10^3 \tev}. 
Only decays with transverse decay length $x_\T > 0.1$~mm are included in the plot.
} 
\label{f.detector}
\end{figure*}
%

\subsection{Constraints and LHC Discovery Prospects} 
\label{s.EightRegions}
This section will discuss the current experimental constraints on our model as well as future discovery prospects. 
For clarity, we classify the parameter space of our model into eight different regions based on the final states as shown 
in \fref{searches}: 
\begin{figure*}[t]
\centering
\begin{tabular}{cc}
\includegraphics[width=2.5in]{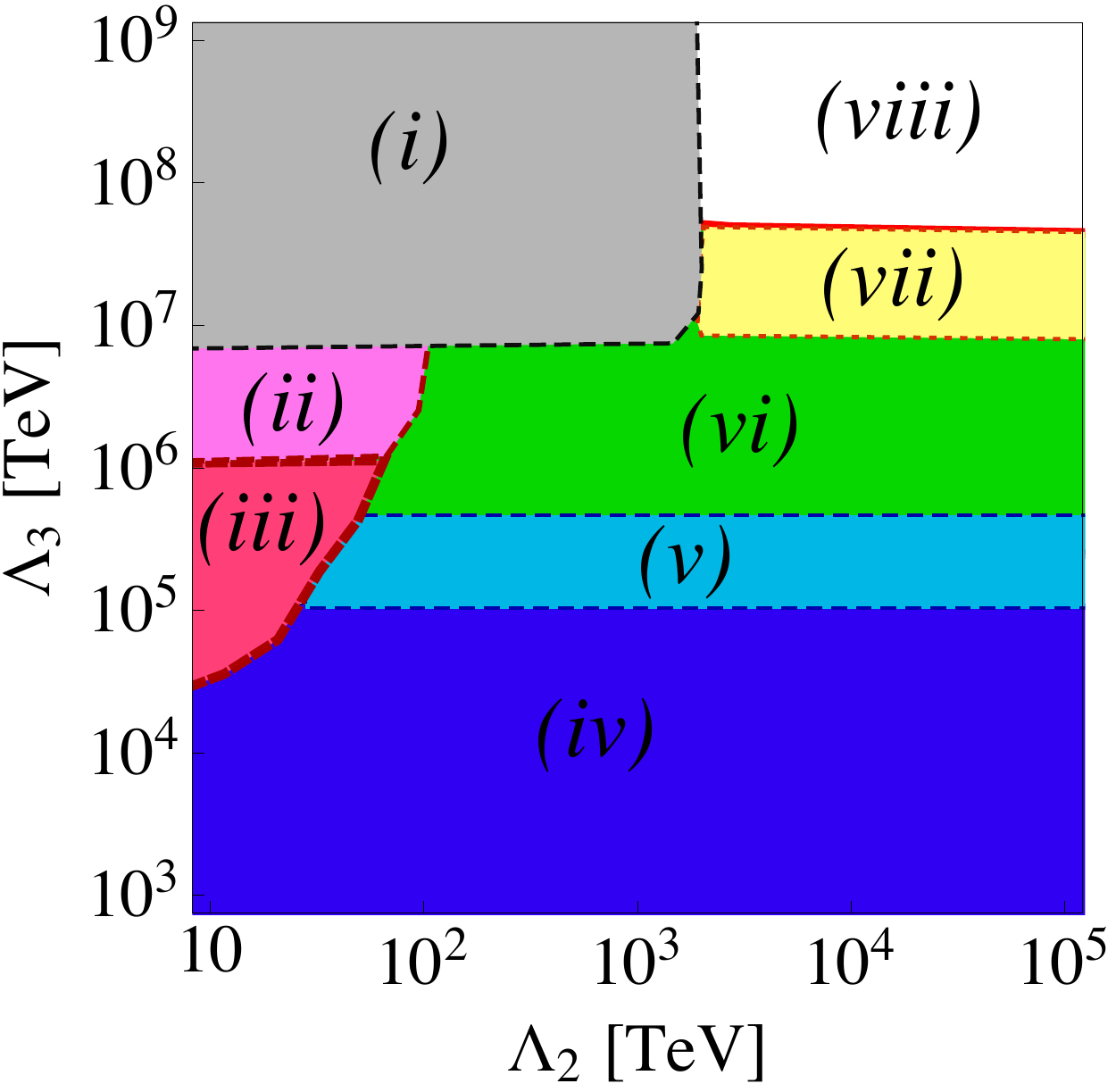}
&\includegraphics[width=3.0in]{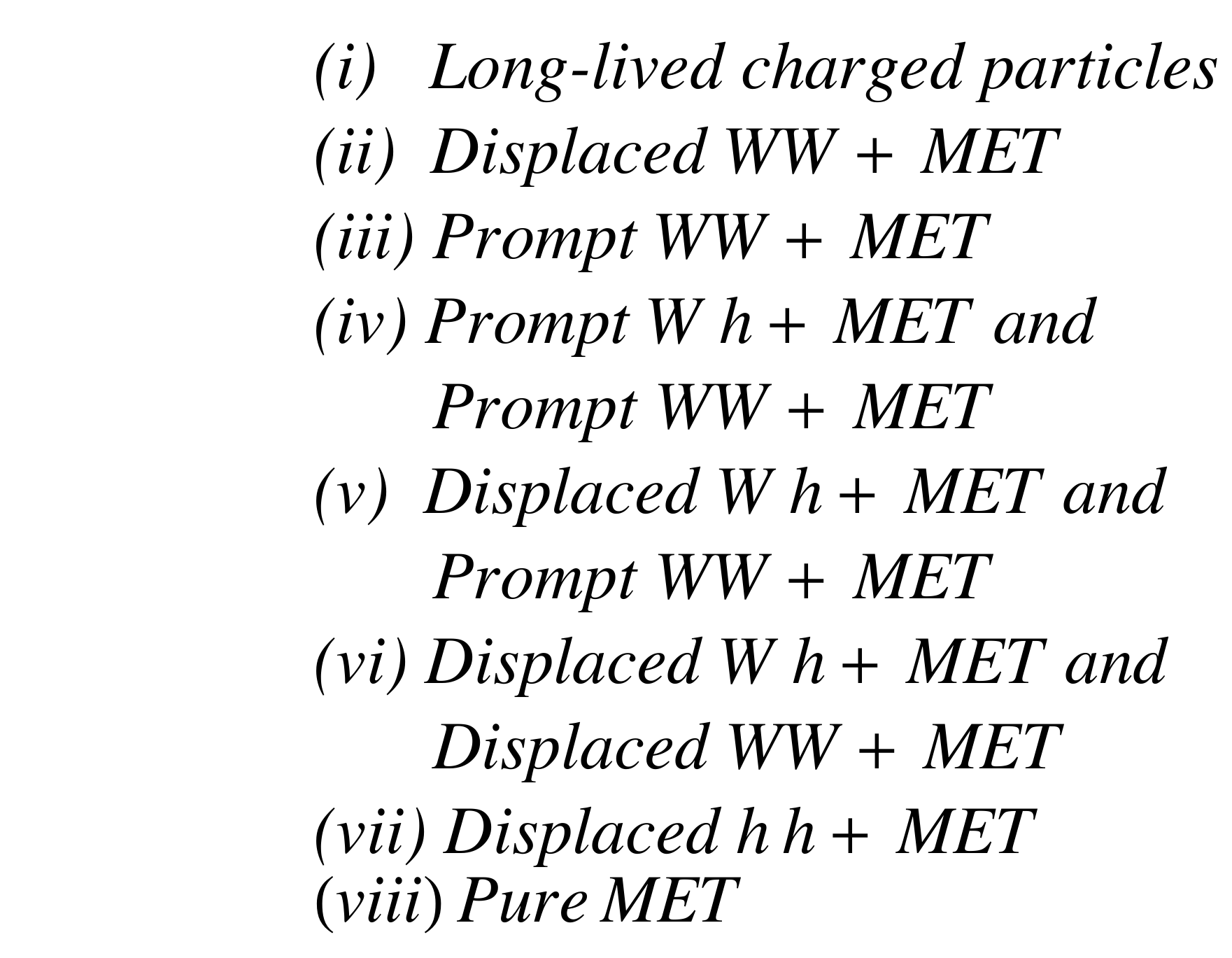} 
\end{tabular} 
\caption{The $\Lambda_2$-$\Lambda_3$ parameter space classified according to the most dominant final states of the $\omega$ decays 
at the LHC for $m_{\omega} = 140 \gev$. The boundary between two given regions is defined as points where the two corresponding 
final states become comparable to each other.
}
\label{f.searches}
\end{figure*} 
%

\begin{center}
{\bf \emph{Region~(i):}} Long-lived charged particles
\end{center}
Here, the charged fermion $\omega_\pm$ has lifetimes exceeding \EQ{50\>}cm. Current LHC searches for long-lived massive charged 
particles are sensitive to this scenario, triggered by activity in the muon tracker from long-lived charged particles or by the large missing energy utilizing only the calorimetric activity. The most constraining of these long-lived charged particle searches is by the LHC experiments  \cite{LongLivedMain}  at $\sqrt{s} = 7 \tev$.  In particular, for lifetimes long enough for the $\omega_\pm$ to reach the muon system of the detector, the parameter space is tightly constrained. For example, in the ATLAS search, the signal efficiency times acceptance for a $\co(100 \gev)$ long-lived charged particles that pass through the detector  is $\sim 20 \%$.  From our elementary detector simulation shown in \fref{LLcharged} for two representative points in the region of parameter space of interest, we estimate that the benchmark point of $m_{\omega} = 140 \gev$ yields $\simgt \co(1)$ events for $\Lambda_3 \simgt 2 \times 10^7 \tev$. Another ATLAS search \cite{ATLAS:LLChargino} has looked for long-lived charginos with disappearing track signatures. This search is sensitive to charged particles that decay in the outermost parts of the inner detector, i.e., with decay lengths $\sim 50$ cm. Combined with other long-lived charged particle searches \cite{LongLived}, we conservatively conclude that our benchmark point 
\EQ{m_{\omega} = 140 \gev} is excluded in \emph{Region (i)}, although a more thorough detector simulation is warranted. 

\begin{figure*}[t]
\centering
\begin{tabular}{cc}
\includegraphics[width=3.2in]{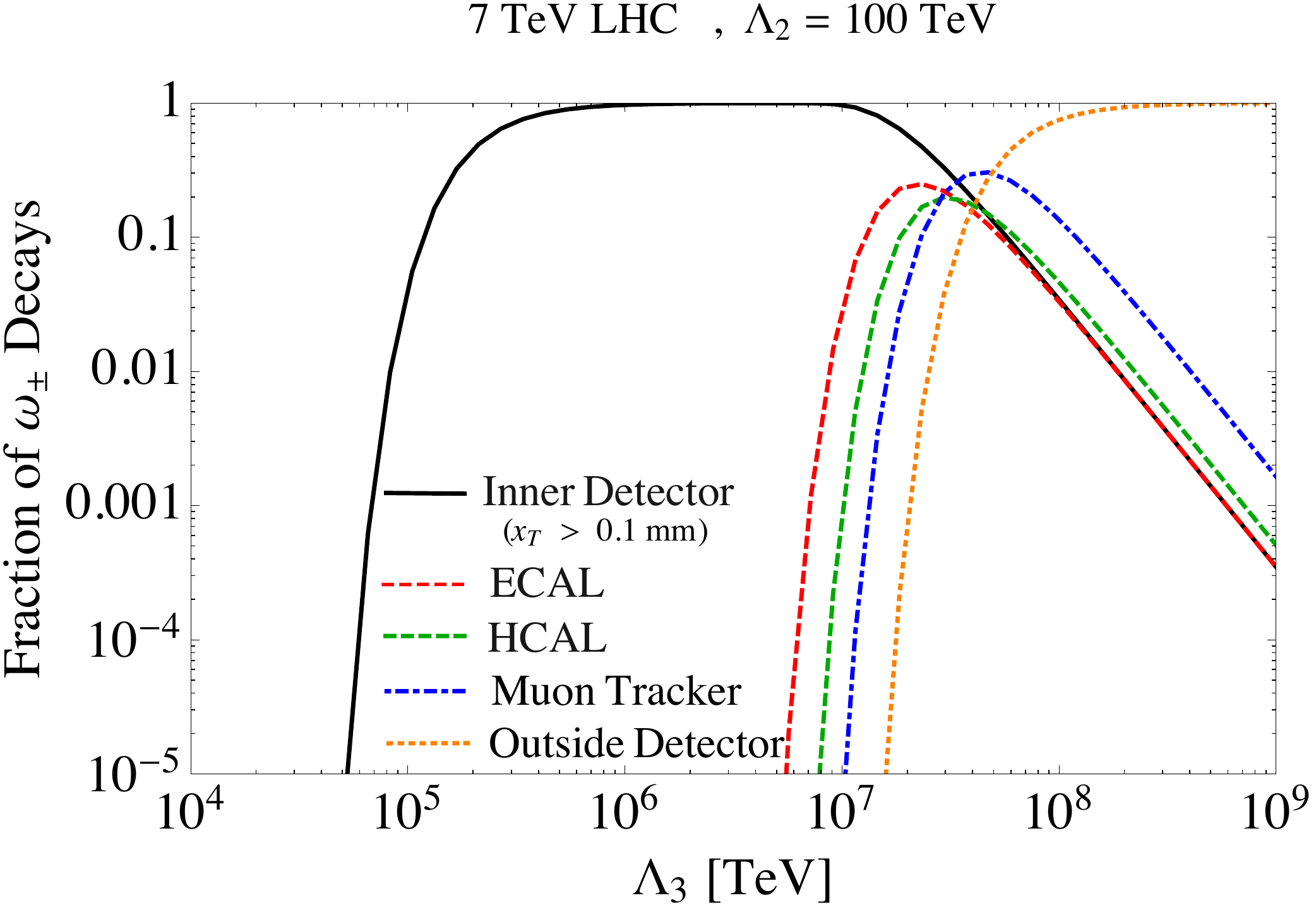}
&\includegraphics[width=3.2in]{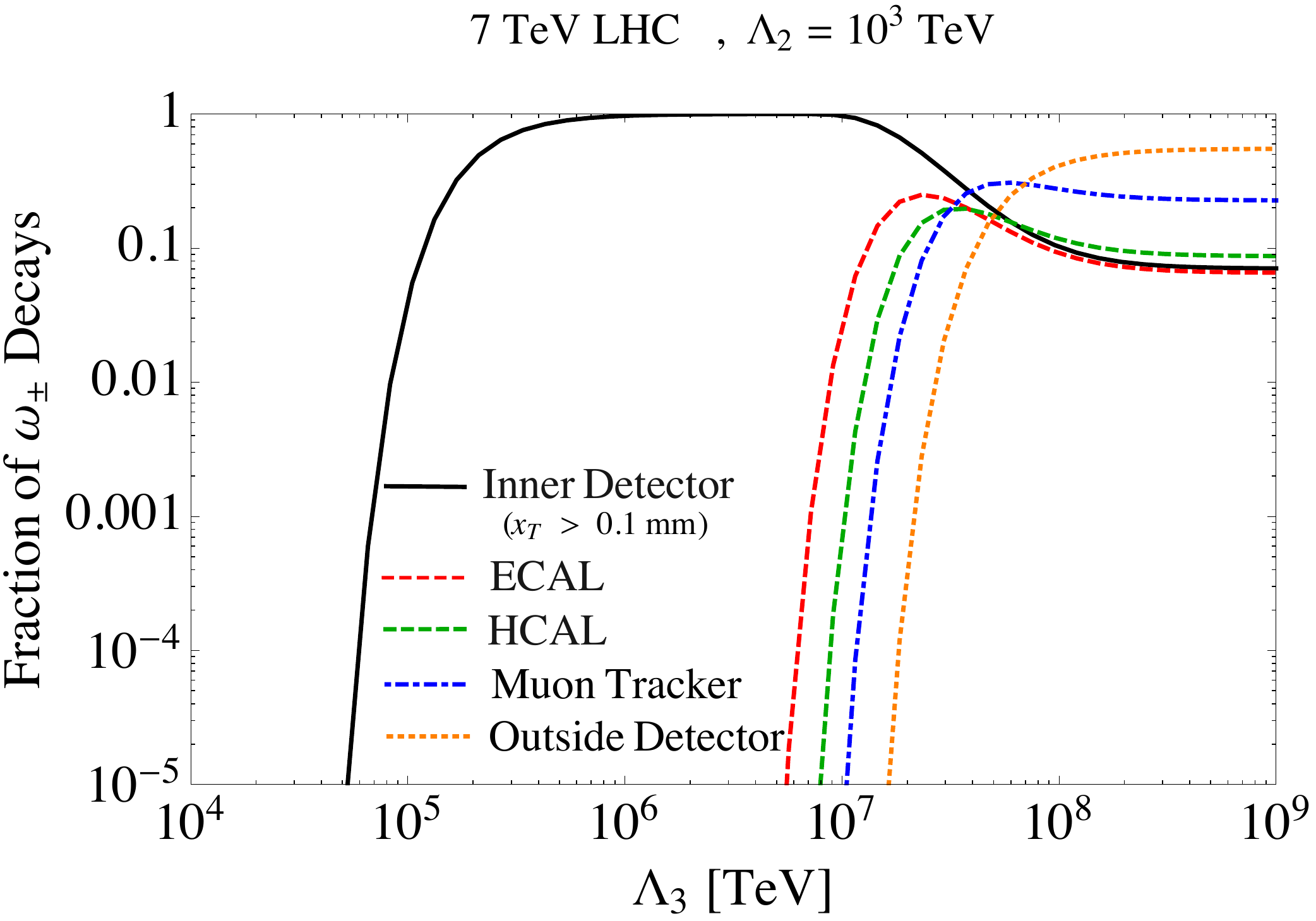}
\\
(a) & (b)
\end{tabular} 
\caption{
Fraction of $\omega_\pm$ decaying inside the various components of the ATLAS detector for 
\EQ{m_\omega = 140\gev} for (a) $\Lambda_2 = 100 \tev$ and (b)  $\Lambda_2 = 10^3 \tev$. 
Only decays with transverse decay length $x_\T > 0.1$~mm are included in the plot.
} 
\label{f.LLcharged}
\end{figure*}
%

\begin{center}
{\bf \emph{Region~(ii):}} Displaced \EQ{WW + \MET} (diagrams in~\fref{FeynWW})
\end{center}
This region of the parameter space is characterized by the following two features:
\begin{itemize} 
\item{$\omega_\pm$ is the NLSP. $\omega_0$ and $\omega_\mp'$ promptly decay to $\omega_\pm$ through weak interactions with 
the decay products from $W^{\pm*}$ being too soft to be detected.} 
\item{$\omega_\pm$ is long-lived, 
but unlike \emph{Region~(i)}, its decay length is in the range between $\sim 1$~mm and $\sim 50$~cm. Below $50$~cm, direct detection of  $\omega_\pm$ at the LHC  rapidly becomes challenging due to an insufficient number of hits in the inner detector.}
\end{itemize}
Therefore, the only potentially observable final state is the \EQ{W^+ W^- + \MET} from the long-lived $\omega_\pm$ decays, 
where both $W^\pm$ are hard and displaced from the primary vertex. 
The hadronic decays of the displaced $W^\pm$ are difficult to observe due to large QCD background. 
For the leptonic decays of the $W^\pm$, while triggers for \EQ{\ell + \MET} already exist at the LHC,
we cannot reconstruct the secondary vertex from which the lepton originates. 
However, searches for \emph{kinked tracks} could be a possible mode for discovery in this region of parameter space.  
\begin{figure*}[t] 
\centering
\includegraphics[width=0.9\linewidth]{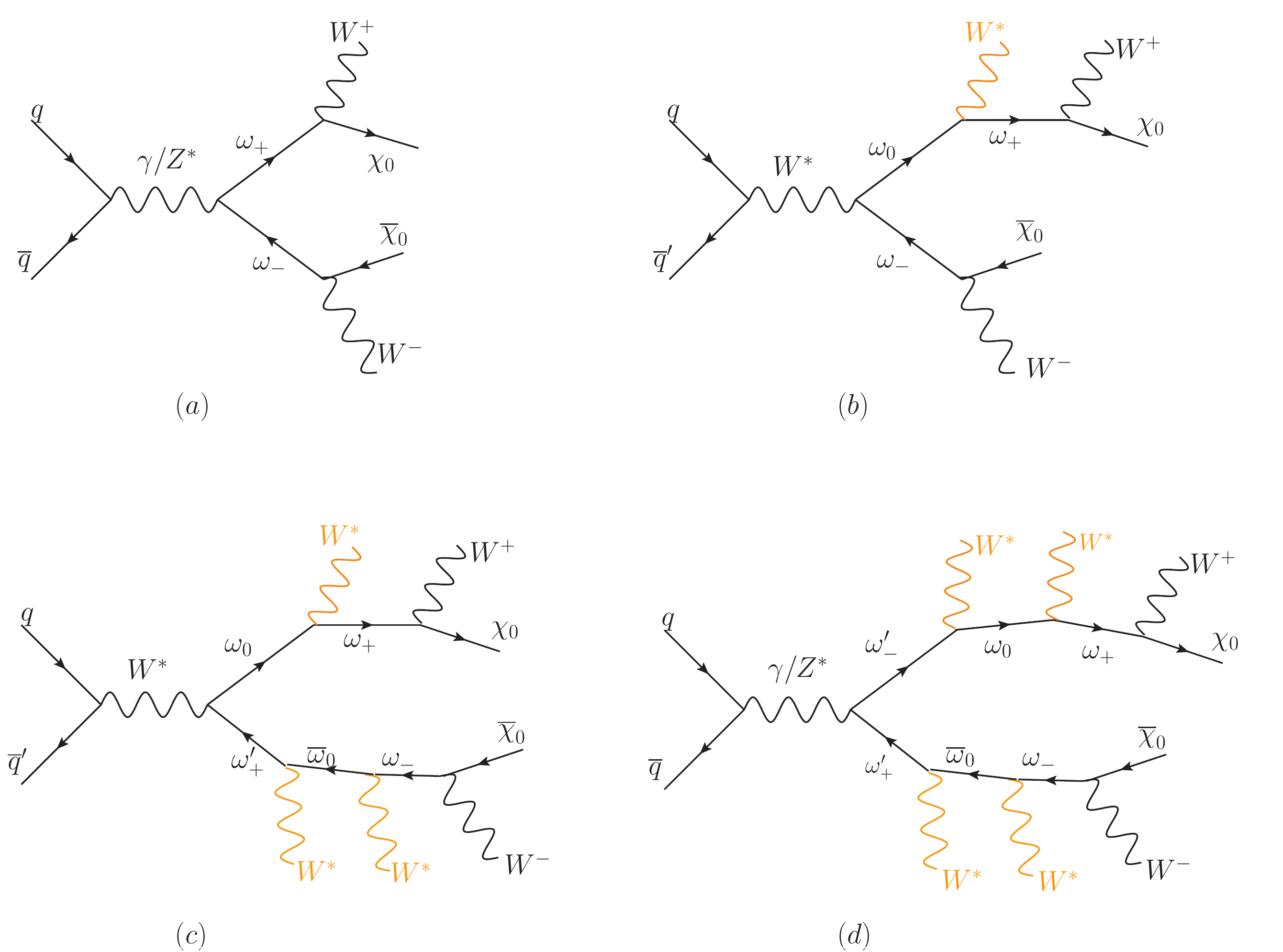}
\caption{Feynman diagrams leading to $W^+ W^- + \MET$ final states in \emph{Regions (ii)} and \emph{(iii)} of the parameter space, 
where $\omega_\pm$ is the NLSP\@. Particles too soft to be detected at the LHC are shown in orange. 
Observe that, in each diagram, the on-shell $W$-pair have opposite signs.} 
\label{f.FeynWW}
\end{figure*}
%

\begin{center}
{\bf \emph{Region~(iii):}} Prompt \EQ{WW + \MET} (diagrams in~\fref{FeynWW})
\end{center}
As in \emph{Region~(ii)}, $\omega_\pm$ is the NLSP here, and 
the only relevant final state is \EQ{W^+ W^- + \MET}, with other decay products from $W^{\pm*}$ being too soft to be observed. 
Unlike \emph{Region~(ii)}, however, $\omega_\pm$ decays promptly, with lifetime $\simlt 1$~mm.
Ignoring the soft decay products, we observe that the Feynman diagrams shown in~\fref{FeynWW} are topologically similar to 
chargino/neutralino pair production in supersymmetric models. 
We have checked that our model is not excluded by opposite-sign dilepton searches~\cite{OSdilepton, SSOSdilepton} 
since they usually require hadronic energy or much larger missing energy compared to our signal. The $m_{T_2}$ searches in \cite{OSdilepton} considered charginos decaying to sleptons which, in turn, give the dilepton signature. It is not immediately clear if this search is applicable to our scenario, though an optimized $m_{T_2}$ search could hold discovery prospects in this region. The trilepton searches~\cite{trilepton} are moot for our scenario, 
since we have at most only two hard leptons. 
Finally, notice that the $\gu{1}_\X$ symmetry implies that the final, on-shell $W$ pair must come in \emph{opposite-sign} 
(see~\fref{FeynWW}). Therefore, this region of parameter space is also safe from same-sign dilepton 
searches~\cite{SSOSdilepton, SSdilepton}. 

The \EQ{W^+ W^- + \MET} production in this region will also manifest as an excess in the SM $WW$ searches~\cite{WW:SM}.
Such an excess can actually improve the agreement with the experiment compared to SM contribution alone, 
as demonstrated in~\cite{WWanomaly} for the case of $WW$ production from $\sim 110 \gev$ charginos decaying to $W$s and gravitinos. 
The analysis in \cite{WWanomaly} also shows that, having similar kinematic features to the SM WW production, 
our \EQ{W^+ W^- + \MET} are unlikely to contaminate the \EQ{h \to WW} searches, unlike the new physics of the type discussed 
in \cite{WWh:fake}, for example.
Therefore, while the uncertainties in the current measurement are too large to make definitive statements, 
the $2\ell+\MET$ searches could be an interesting path to discovery should our model prove to be correct. 

\begin{figure*}[h]
\centering
\begin{tabular}{cc}
\includegraphics[width=2.9in]{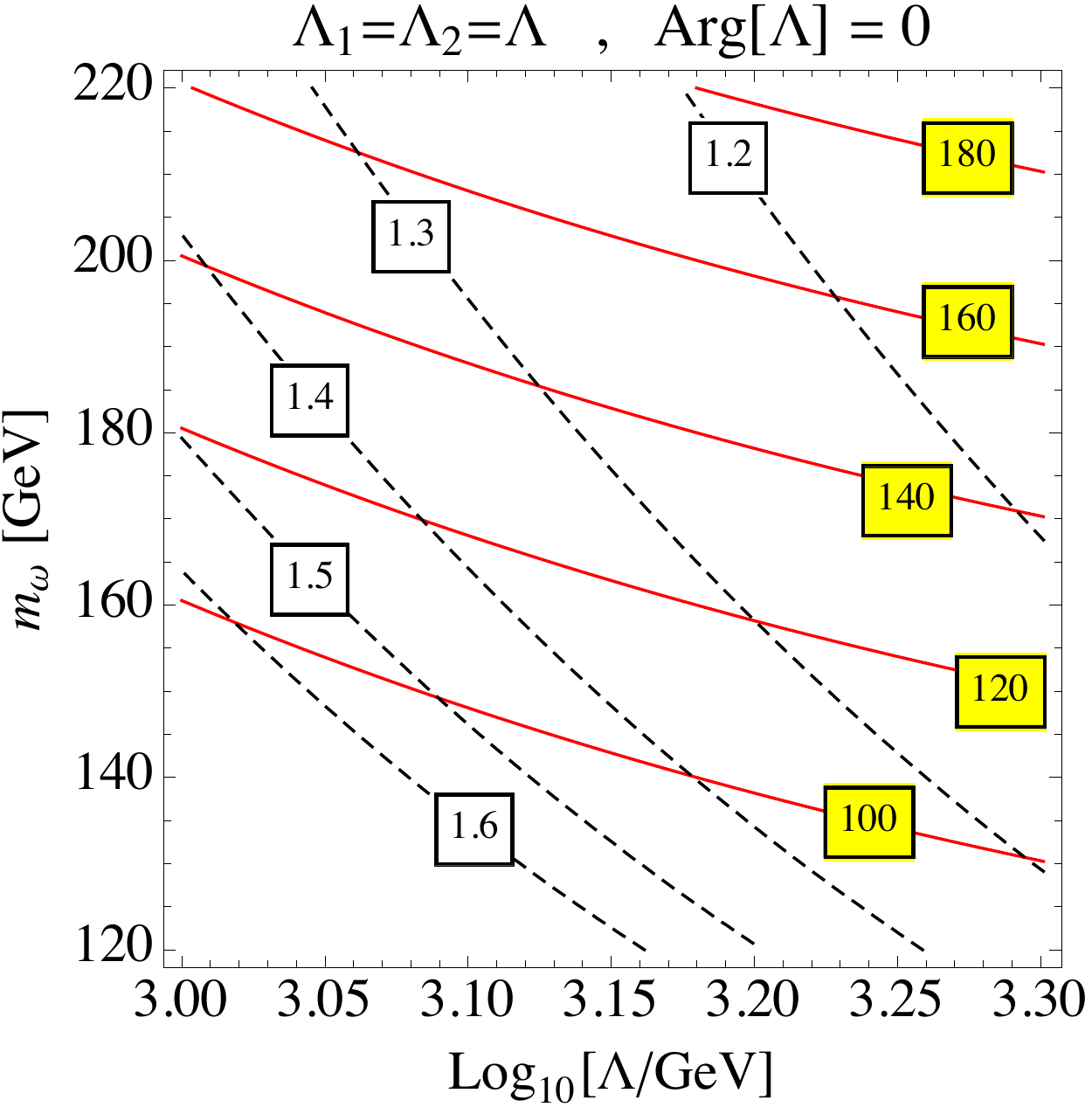}
& \quad
\includegraphics[width=2.9in]{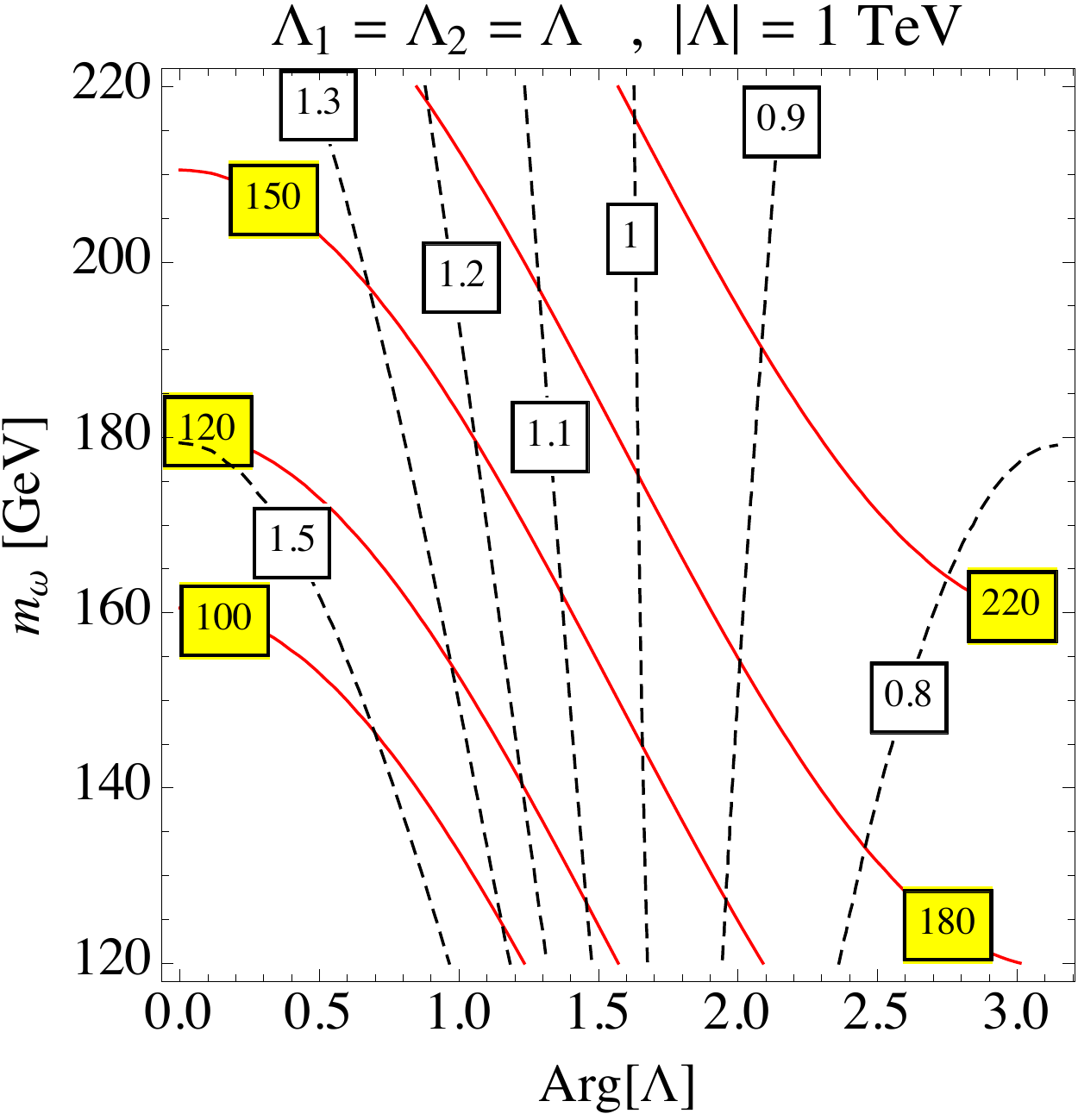} 
\\
(a) & (b)
\end{tabular} 
\caption{Contour plots of the diphoton rates $R_{\gamma \gamma}$ (dashed black lines, white boxes) 
and the masses of lightest $\omega$ fermion, $m_{\omega_0}$ (solid red lines, yellow boxes) 
are shown for \EQ{\Lambda_3 \gg \Lambda_1, \Lambda_2}, 
assuming \EQ{\Lambda_1 = \Lambda_2}. 
In plot~(a), contour lines are plotted in $\Lambda_1$-$m_\omega$ plane for real and positive $\Lambda_{1,2}$. 
In plot~(b), the magnitude of the scales $\Lambda_1$ and $\Lambda_2$ is fixed at $1 \tev$ while their complex phase 
$\text{Arg}[\Lambda]$ is varied.}
\label{f.Diphoton}
\end{figure*} 

The operator~$\co_5^{(1)}$, which is mostly inconsequential and has thus been ignored so far, can have an interesting effect 
in \emph{Region~(iii)}.
Although outside the plot in \fref{searches}, 
the Higgs diphoton rate, \EQ{p \, p \to ( h \to \gamma \gamma) + X}, can be significantly modified  
if \EQ{\Lambda_1 \sim \Lambda_2 \sim 1 \tev}. 
Including the contributions of the top quark, $W^\pm$ and the new charged fermions, $\omega_\pm$ and $\omega_\mp'$, 
the decay width of the Higgs to diphoton is given by 
\begin{equation}
\Gamma ( h \to \gamma \gamma ) = \frac{\alpha_\text{EM}^2 \, G_\text{F} \, m_h^3}{128 \sqrt{2} \, \pi^3} \, 
\left| 
A_1^h(\tau_\W) 
+ \frac{4}{3} A_{1/2}^h (\tau_t) 
+ \sum_{f=\omega, \, \omega'}  g_{h f_+ f_-}^\PD \frac{v}{m_{f_\pm}} A_{1/2}^h (\tau_{f_\pm}) 
\right|^2
,
\end{equation}
where $\tau_i = m_h^2/(4 m_i^2)$, the functions $A_{1/2}^h, \, A_1^h$ are defined in \cite{Djouadi2}, 
and the couplings $g_{h f_+ f_-}$ for the charged $\omega$ fermions induced by the operators $\co_5^{(1,2)}$ are given by 
\begin{equation}
g_{h \omega_+ \omega_-}^\PD 
= -\frac{v}{\Lambda_1} - \frac{v}{\Lambda_2} 
\,,\quad 
g_{h \omega_+' \omega_-'}^\PD 
= - \frac{v}{\Lambda_1} + \frac{v}{\Lambda_2} 
\,.
\end{equation}
For simplicity, we consider the UV completion~\erefn{UVcompletion} so that \EQ{\Lambda_1 = \Lambda_2 \equiv \Lambda}, 
where $\Lambda$ in general is complex. The diphoton rate $R_{\gamma \gamma}$ is plotted in \fref{Diphoton}, where%
\begin{align}
R_{\gamma \gamma}
\equiv
\frac{\sigma (p p \to h + X) \times Br(h \to \gamma \gamma)}{\sigma (p p \to h+X) \times Br(h \to \gamma \gamma) \bigr|_\text{SM}}
\,.
\end{align}
We find that the diphoton rates are maximized if $\Lambda$ is close to real and positive.

\begin{center}
{\bf \emph{Region~(iv):}} Prompt \EQ{Wh + \MET}, Prompt \EQ{WW + \MET} (diagrams in~\fref{FeynWH})
\end{center}
In this region, all the $\omega$ fermions decay promptly to $\chi_0$ in a single step. 
\fref{FeynWH} shows two possible final states of interest at the LHC, \EQ{W h + \MET} and \EQ{W^+ W^-+ \MET}. 
Undoubtedly, the strongest constraint for this final state comes from the associated Higgs production searches in the 
$b \overline{b}$ channel, \EQ{p p \to W + (h \to b \overline{b})}. 
The $95\%$~CL limits on \EQ{\sigma(p p \to V h) \times BR( h \to b \overline{b})} (\EQ{V= W}, $Z$) 
by ATLAS and CMS for $\sim 125 \gev$ Higgs are $1.8$ and $2.5$ times the SM value, respectively~\cite{hbb}. 
Most of our signal events pass the kinematic cuts used in these searches for \EQ{m_{\omega} \simlt 200 \gev}, 
thus comfortably excluding the benchmark point $m_\omega = 140\gev$. 
This situation is similar to a supersymmetric scenario with \EQ{\mu \gg M_1, M_2}, 
where the NLSP neutralino can decay to Higgs and the LSP, and the chargino decays to $W$ and the LSP~\cite{Wh:MSSM, Wh:Prompt}. 

\begin{center}
{\bf \emph{Region~(v):}} Displaced \EQ{Wh + \MET}, Prompt \EQ{WW + \MET} (relevant diagrams in~\fref{FeynWH})
\end{center}
In this region, the neutral fermion $\omega_0$ is long-lived with lifetime in the range \EQ{1\>}mm--\EQ{1\>}cm. 
The charged states, $\omega_\pm$ and $\omega_\mp'$, still decay promptly with \EQ{\simlt 1\>}mm. 
This difference stems from the proximity of $m_h$ to $m_\omega$, 
which leads to a slightly more suppressed phase space for \EQ{\omega_0 \to h \chi_0} than \EQ{\omega_+ \to W^+ \chi_0}, 
while the interaction strengths behind these decays are parametrically the same.

Since the lifetime of $\omega_0$ is \EQ{\simlt 1\>}cm, it typically decays to \EQ{h + \MET} before reaching the inner detector, 
so the actual final states to be observed are the $B$ hadrons from the $h$.%
\footnote{A fraction of $\omega_0$ will decay inside the inner detector, for which 
we refer the reader to section on \emph{Decays inside the inner detector or before} in \emph{Region~(vi)}.}
Notice, however, that these $B$ hadrons originate from a vertex that itself is already displaced, as the decay 
\EQ{\omega_0 \to h + \MET} has lifetime \EQ{\simgt 1\>}mm.  
This significantly degrades the efficiency of standard $b$-tagging algorithms~\cite{b-tagging} that use a ``signed impact parameter'' 
as a discriminating variable with the positive sign being preferred. 
As illustrated in \fref{displaced}, the sign is determined by the angle between the decay length vector 
and the jet-axis, where the sign is taken to be positive if this angle is less than $90^\circ$. As shown in \fref{displaced}(b), 
a significant fraction of $b$ jets originating from displaced $\omega_0$ decays will give negatively signed impact 
parameters, resulting in a much reduced efficiency of tagging the $b$ quarks in our signals,
although a detailed simulation of this effect is beyond the scope of a theoretical paper. 
Let us conclude that, unlike \emph{Region~(iv)}, we expect that the \EQ{W h} searches should not be very constraining, 
and, turning this around, the observation of events with negative impact parameters should be regarded as the opportunity 
to probe this scenario.

\begin{figure*}[t] 
\centering
\includegraphics[width=0.7\linewidth]{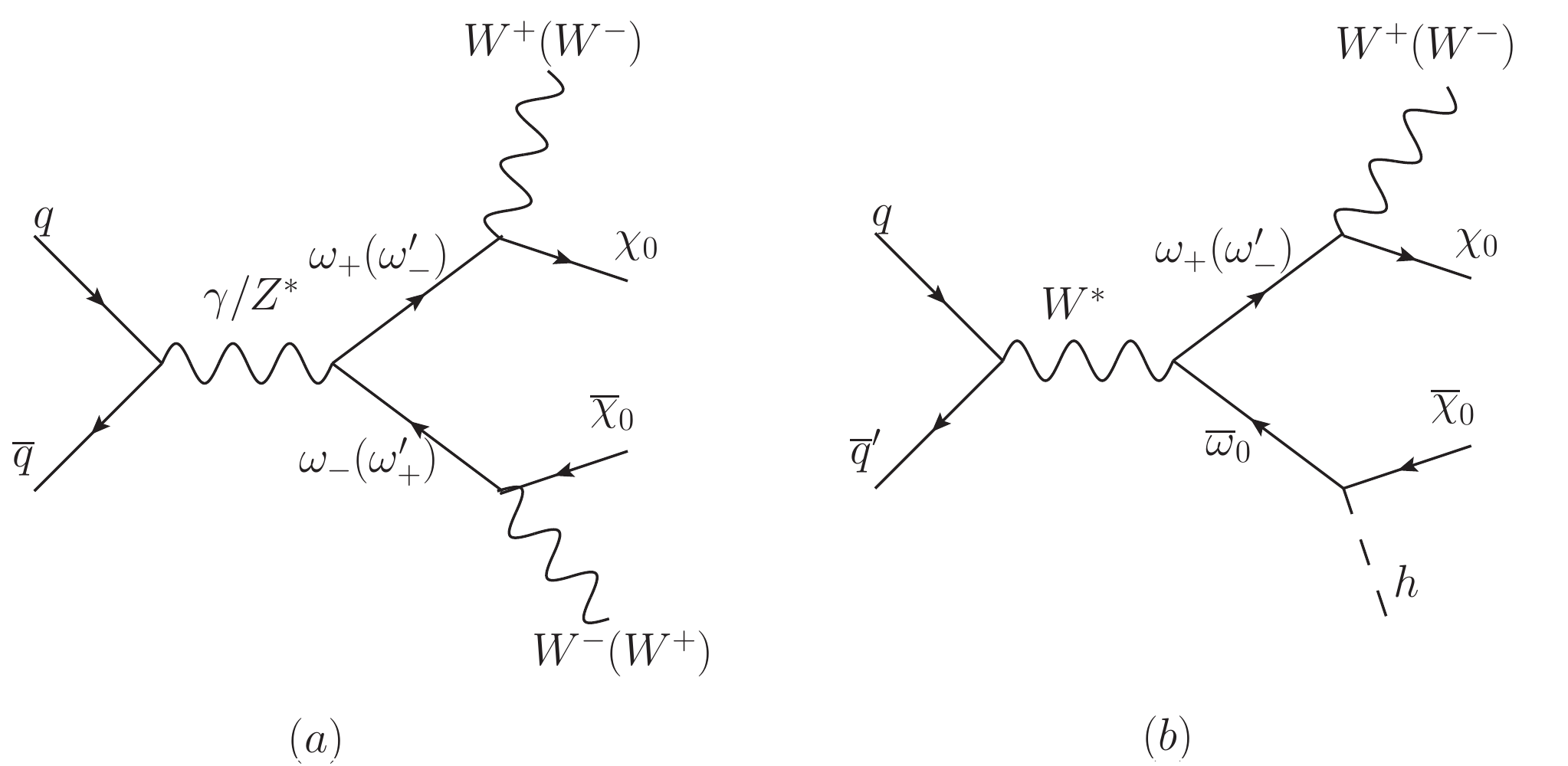}
\caption{Feynman diagrams leading to $W^+ W^- + \MET$ and \EQ{W^\pm h + \MET} final states 
in \emph{Regions (iv), (v)} and \emph{(vi)}, where each $\omega$ fermion predominantly decays to $\chi_0$.} 
\label{f.FeynWH}
\end{figure*} 

Given that the current \EQ{W h} searches do not exclude the parameter space of interest, 
we are left with the \EQ{W^+ W^- + \MET} final state produced from the decays of charged $\omega$ fermions 
with lifetime \EQ{\simgt 1\>}mm (see \fref{FeynWH}(a)). 
This is regarded as prompt decays by ATLAS and CMS, both of which require the transverse impact parameter to be \EQ{|d_0| \simlt 1\>}mm. 
Such prompt $WW$ final states have already been discussed for \emph{Region~(iii)}.
\begin{figure*}[t]
\centering
\begin{tabular}{cc}
\includegraphics[width=2.8in]{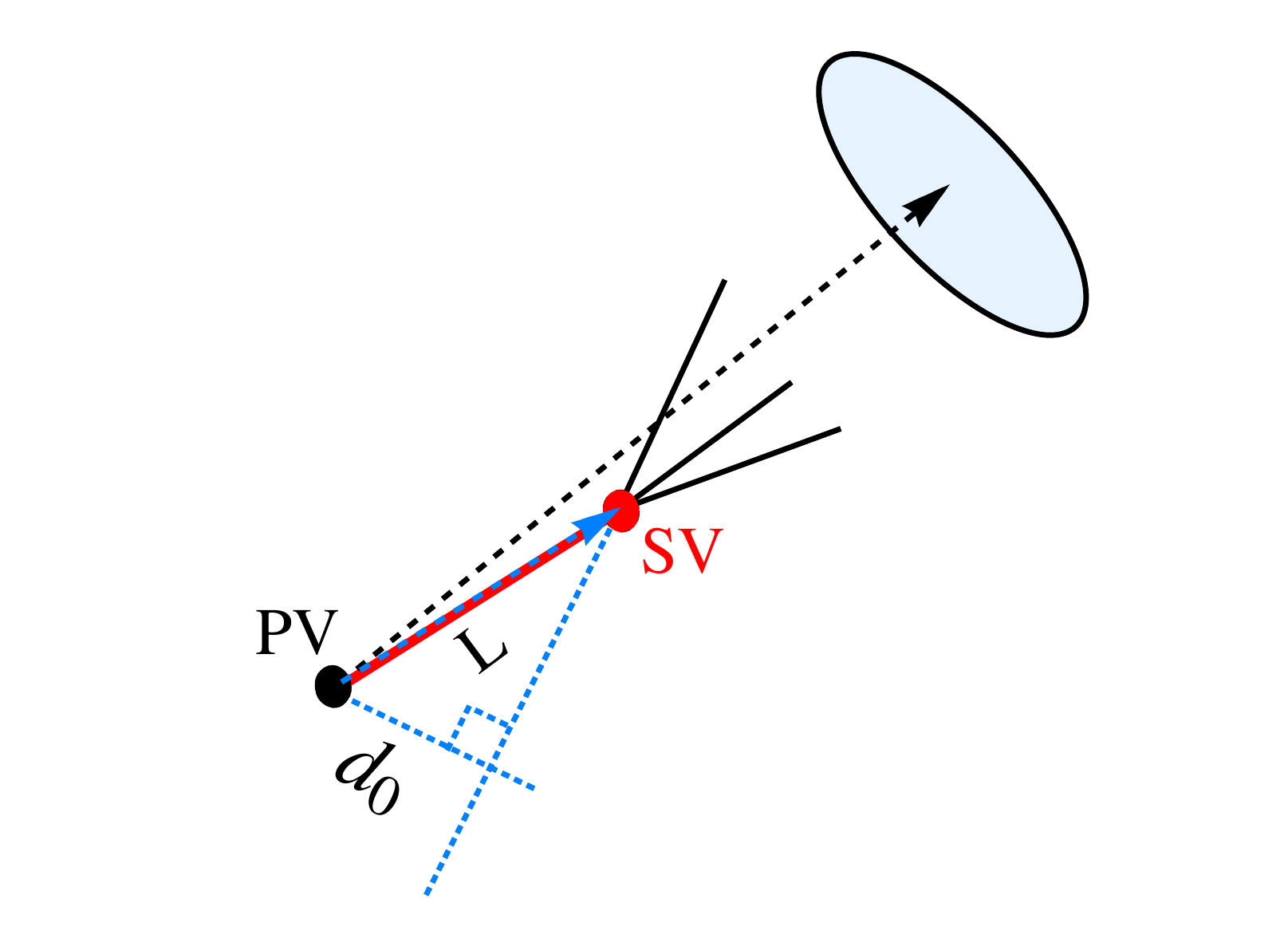}
&\includegraphics[width=2.8in]{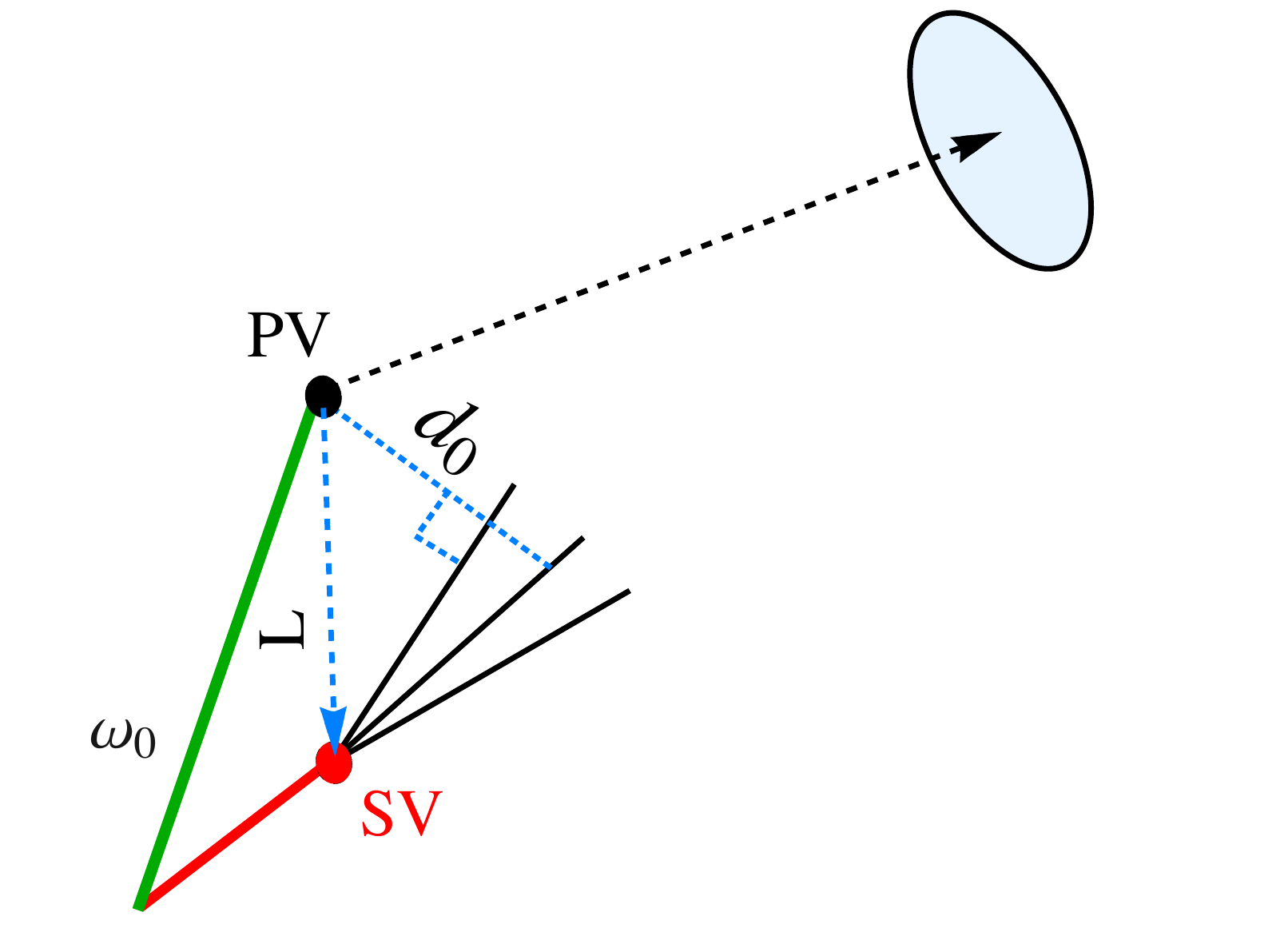}
\\
(a) & (b)
\end{tabular} 
\caption{
Diagrams show $b$-tagging using secondary vertex (SV) in processes where (a) the $b$-quark (shown in red) 
is emitted from primary vertex (PV), 
and (b) the $B$-hadron (shown in red) is emitted from the decay of a long-lived particle such as $\omega_0$ (shown in green). 
In both cases, the jet axis is represented by dashed black arrow, and the decay length vector $L$ by the dashed blue arrow. Note that the decay length in diagram (a) coincides with the $B$-hadron track. 
In both diagrams (a) and (b), $d_0$ is the impact parameter for a particular track. 
Diagram (a) has positive signed impact parameter, while (b) has a negative signed impact parameter (refer to text for details).
} 
\label{f.displaced}
\end{figure*} 
%

\begin{center}
{\bf \emph{Region~(vi):}} Displaced \EQ{Wh + \MET}, Displaced \EQ{WW + \MET} (diagrams in~\fref{FeynWH})
\end{center}
This region displays a rich assortment of exotic signals, since both the charged and neutral $\omega$ decay through displaced vertices 
to $\chi_0$, as in~\fref{FeynWH}.
The lifetimes of the charged $\omega$ states are in the range $1$--\EQ{50\>}mm, 
so they still decay before reaching the inner detector, 
rendering them safe from the searches for long-lived charged particles, as discussed for \emph{Region~(i)}.  
On the other hand, the neutral state, $\omega_0$, can have lifetimes as long as a few meters 
and can thus decay to Higgs and missing energy anywhere inside the detector. 
The relevant final states are displaced \EQ{W W + \MET} and \EQ{W h + \MET}, where the $W$ is displaced by at most \EQ{\sim 5\>}cm, 
while the $h$ can appear anywhere in the \emph{middle of the detector}. 
The fraction of $\omega_0$ decays in various parts of the detector are plotted in \fref{detector}(b).

As discussed for \emph{Region~(ii)}, 
the displaced $W$ are difficult to search for due to the QCD background in the hadronic decay channel and 
the inability to reconstruct the secondary vertex in the leptonic decay channel. 
Nevertheless, the leptons from displaced $W$ decay can be used for triggering purposes, 
although trigger efficiency is expected to be low since the leptons may not be high-$p_\T$ for our benchmark point.
On the other hand, for the displaced \EQ{W h + \MET} signal,  
the Higgs bosons can be used to reconstruct the corresponding secondary vertex in many cases.
Below we discuss the collider signatures and relevant searches depending on where the $\omega_0$ decays inside the detector: 
\begin{itemize}
\item{{\bf Decays inside the inner detector or before}: 
This section covers the parameter space with $\omega_0$ lifetimes \EQ{\simlt 1\>}m, including the range $1$--\EQ{10\>}mm 
of \emph{Region~(vi)}.  
For $\Lambda_3$ between \EQ{\approx 6 \times 10^5 \tev} and \EQ{\approx 3 \times 10^6 \tev},
$\omega_0$ will decay to \EQ{h + \MET} predominantly inside the inner detector or the beam pipe. 
The Higgs boson thus produced will then decay to a number of final states: 
\begin{itemize}
\item[(a)]{\underline{\EQ{h\to b \overline{b}}}: 
This is the dominant decay mode of the Higgs boson, with \EQ{BR(h \to  b \overline{b}) \approx 58 \%} for \EQ{m_h \approx 125\gev}. 
CDF and D0 conducted searches~\cite{LLNeutral:bb} for long-lived neutral particles decaying to $b\bar{b}$ 
with lifetimes of \EQ{\sim 2\>}cm and \EQ{\sim 2.5}--\EQ{10\>}cm, respectively, 
in the context of hidden valley scenarios~\cite{HiddenValley_Higgs}, where the Higgs decays to a pair of long-lived neutral particles 
$\pi_\text{v}$, which in turn decay to $b \overline{b}$, thus giving rise to \EQ{h \to \pi_\text{v} \pi_\text{v} \to 2b 2\overline{b}}. 
The applicability of these searches to our case is far from clear, however, because our $h$ is not highly boosted like $\pi_\text{v}$, 
which should affect the efficiencies. 
Nonetheless, given that triggers already exist for highly displaced $b$ jets (through high-$p_\T$ muons produced from $B$-hadron decays), 
it would be worthwhile to look into our displaced Higgs signals in the Tevatron data.
}
\item[(b)]{\underline{\EQ{h \to n\>\text{jets} + X} (\EQ{n \geq 2})}: 
Two or more jets in the final states can arise from a number of Higgs decay channels, e.g., 
\EQ{h \to g g}, \EQ{h\to b \overline{b}}, \EQ{h \to W W^* \to \text{4j}} or \EQ{\ell \nu \text{2j}}, 
\EQ{h\to Z Z^* \to 4j} or \EQ{2j + X}. 
If the $\omega_0$ decays to jets (through Higgs) between \EQ{\sim 40\>}cm--\EQ{100\>}cm of the ATLAS inner detector, i.e., 
outside the pixel detector but inside the silicon and transition radiation trackers, 
then the decays would appear as \emph{trackless jets}. 
While currently no searches exist for this signature, it is our understanding that ATLAS has implemented~\cite{HVTrigger} 
a signature-driven trigger for trackless jets in the hidden valley context. 
Such a trigger should also help selecting events in our model. 
}
\item[(c)]{\underline{\EQ{h \to \ell + n\>\text{jets} + X}}: 
Single-lepton triggers may be used to select this channel, where the lepton comes from  
\EQ{\omega_0 \to (h \to WW^* / ZZ^*) + \MET} or \EQ{\omega_\pm / \omega'_\pm \to W^\pm + \MET}. 
In particular, the final state \EQ{h \to W W^* \to \mu \nu \text{2j}} is
similar to signatures of R-parity violating supersymmetric models where the long-lived neutralino decays to muons and jets with 
a displaced vertex. Hence, the ATLAS searches for such neutralinos~\cite{LLNeutral:Muon+Jets}, where 
the trigger requires a muon to have \EQ{p_\T > 50 \gev}, should apply to our case with a similar sensitivity.  
}
\item[(d)]{\underline{\EQ{h \to n \ell + X} (\EQ{n \geq 2})}: 
Single-lepton or double-lepton triggers can be used to select such events.
The multi-lepton final states can arise from \EQ{\omega_0 \to (h \to WW^*/ZZ^*) + \MET}. 
A search for long-lived particles decaying to \EQ{\ell \ell + \MET} was conducted at D0~\cite{LLNeutral:mumu, LLNeutral:diphoton}, 
and we have checked that our benchmark point \EQ{m_\omega = 140\gev} is safe from this search due to weak exclusions at the Tevatron.
The CMS experiment has searched for neutral long-lived resonances decaying to dileptons~\cite{LLNeutral:dilepton}. 
In our model, such displaced dilepton resonance is provided by the $Z$ from \EQ{\omega_0 \to (h \to Z Z^*) + \MET}. 
Our model, however, is safe from this CMS search because of a decreased sensitivity near the $Z$ resonance as well as 
a large lepton $p_\T$ cuts used in the search.  
Searches for displaced and highly boosted \emph{muon jets}~\cite{LLNeutral:muonjets} are not directly relevant to our scenario, 
as our muons will not be very boosted. 
}
\item[(e)]{\underline{\EQ{h \to \gamma  \gamma} or \EQ{\gamma Z}}: 
Another smoking gun signature of this region of parameter space is the displaced \EQ{\gamma\gamma} or \EQ{\gamma Z} resonance 
that reconstruct to the Higgs mass. 
Although there are existing searches for displaced single or two photons~\cite{LLNeutral:GMSB:PhotonMET, LLNeutral:diphoton}, 
none of them seek a resonance from a diphoton pair, as they are primarily motivated by gauge mediated supersymmetric models. 
}
\end{itemize}
}
\item{{\bf Decays inside the calorimeter}: 
There are currently no searches for long-lived particles decaying inside the calorimeter, except for stopped particles searches \cite{stopped}, which are not directly relevant for our scenario. 
ATLAS, however, has implemented triggers for long-lived neutral particles decaying inside the HCAL~\cite{HVTrigger}.
These triggers are motivated by hidden valley models but can be just as sensitive, if not more, to our displaced Higgs. 
Events are triggered by the following two signatures designed to characterize decays inside HCAL: 
(a) The decay products are confined to a small region inside the calorimeter, and 
(b) The ratio of energy deposition inside the ECAL to that in HCAL is very small. 
Depending on the luminosity, the benchmark point of our model can be sensitive to this search channel 
for \EQ{\Lambda_3 \sim 10^7 \tev} and \EQ{\Lambda_2 \simgt 100 \tev}. 
}
\item{{\bf Decays inside the muon spectrometer}: 
\fref{detector}(b) shows that a significant number of decays will occur in the muon spectrometer 
in this region of the parameter space. 
ATLAS has carried out a search~\cite{HV:MuonTracker} 
for hidden valley scenarios in which the long-lived neutral particles $\pi_\text{v}$ decay inside the muon tracker.
There is, however, an important difference between the $\pi_\text{v}$ and our displaced particle, $h$. 
Namely, the $\pi_\text{v}$ masses considered by the ATLAS search are \EQ{\sim 20}--\EQ{40\gev}, while the $h$ mass is \EQ{\approx 125\gev}.
This implies that our displaced $h$ will be far less boosted compared to the $\pi_\text{v}$. 
Consequently, the trigger efficiencies would be lower for our case, 
as some fraction of the hits in the muon spectrometer due to our $h$ would be associated to a wrong bunch crossing.
A realistic estimation of the trigger efficiencies as well as the sensitivity of the ATLAS search for our scenario would 
require a detailed detector simulation, which is beyond the scope of a theory paper.
}
\end{itemize}
\begin{figure*}[t]
\centering
\qquad \includegraphics[width=0.8\linewidth]{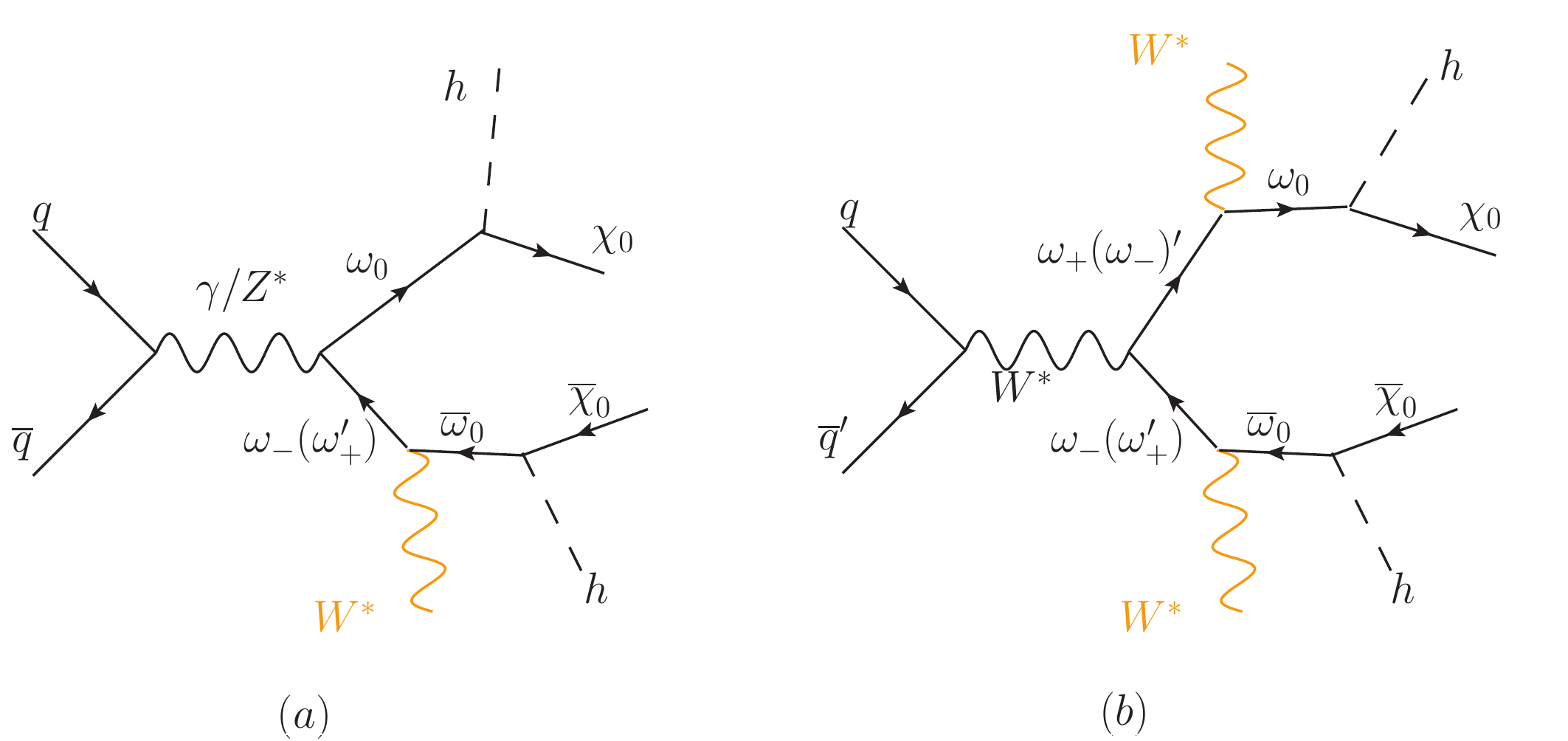}
\caption{Feynman diagrams leading to $hh + \MET$ final states in \emph{Region~(vii)} of the parameter space, 
where the charged $\omega$ fermions predominantly decay to $\omega_0$. 
Particles too soft to be detected at the LHC are shown in orange.} 
\label{f.FeynHH}
\end{figure*} 
%

\begin{center}
{\bf \emph{Region~(vii):}} Displaced \EQ{hh + \MET} (diagrams in \fref{FeynHH})
\end{center}
Here, the charged $\omega$ states decay predominantly to $\omega_0$ instead of $\chi_0$ with relatively short lifetimes \EQ{\simlt 10\>}cm.
The lifetime of $\omega_0$, on the other hand, can be as long as a few meters. 
As shown in \fref{FeynHH}, the relevant final state is $hh + \MET$, where both Higgs bosons are macroscopically displaced 
from the primary vertex. 
The phenomenology of displaced Higgs has already been discussed in great detail for \emph{Regions~(v)} and~\emph{(vi)}, 
although it should be pointed out that the efficiency for detecting a displaced Higgs signal in this region goes up by a factor of two, 
simply because there are two Higgs bosons.

\begin{center}
{\bf \emph{Region~(viii):}} Pure $\MET$
\end{center}
In this region, the relevant diagrams are given by \fref{FeynHH}, but 
the neutral state $\omega_0$ is so long-lived (\EQ{\simgt 10\>}m) 
that it decays predominantly outside the detector (see \fref{detector}(b)). Similar to dark matter searches using $\MET + X$ ($X=j,\gamma,W,Z$), 
this region can be probed at the LHC using jets, photons or $W/Z$ bosons from the initial state partons. The relevant searches are : $\MET + j$ \cite{DM:jet}, $\MET + \gamma$ \cite{DM:photon} and $\MET + W/Z$ \cite{DM:WZ}. For the $\MET + j$ case, we generated the $\omega \omega + 0j$ and $\omega \omega + 1j$ ($p_\T^j > 80 \gev$) samples in {\tt MadGraph} and showered/hadronized the partonic events in {\tt Pythia}. The partonic cross-sections for the $0$-jet and $1$-jet samples were found to be $5.50$ pb and $0.22$ pb respectively at $7 \tev$.  For the CMS, the efficiencies for passing $\MET > 350 \gev$, jet $p_\T > 110 \gev$ and other cuts was found to be $\sim 0.2\%$ and $\sim 2\%$ for $0$-jet and $1$-jet samples respectively. Even without performing any matching, we obtain at most $\sim 70$ events with a luminosity of $5 \fb^{-1}$ whereas the $95\%$ CL limit on non-SM like events placed by CMS from this search is $158$ events. A similar analysis for the ATLAS experiment and for other final states with photons/$W/Z$ bosons yields the same conclusion that this region is not excluded by $\MET + X$ searches. However, improved sensitivity is expected with $8 \tev$ data.

\section*{Acknowledgment} 
We would like to thank Todd Adams, Can Kilic, Patrick Meade and Harrison Prosper for discussions and comments on our manuscript. We would also like to thank Andrew Askew, David Curtin, Venkatesh Veeraraghavan  and Marc Weinberg for useful discussions. This work is supported by the DOE under grant DE-FG02-97ER41022.



\begin{thebibliography}{99}



\bibitem{Higgs125Combined} 
  G.~Aad {\it et al.}  [ATLAS Collaboration],
  Phys.\ Lett.\ B {\bf 716}, 1 (2012)
  [arXiv:1207.7214 [hep-ex]];
  
  S.~Chatrchyan {\it et al.}  [CMS Collaboration],
  Phys.\ Lett.\ B {\bf 716}, 30 (2012)
  [arXiv:1207.7235 [hep-ex]].
  
 \bibitem{Higgs125gaga} 
  [ATLAS Collaboration],
  ATLAS-CONF-2012-091;
   
  [CMS Collaboration],
  CMS-PAS-HIG-12-015.



\bibitem{HiddenValley_Original} 
  M.~J.~Strassler and K.~M.~Zurek,
  Phys.\ Lett.\ B {\bf 651}, 374 (2007)
  [hep-ph/0604261].
  
\bibitem{HiddenValley_Higgs} 
  M.~J.~Strassler and K.~M.~Zurek,
  Phys.\ Lett.\ B {\bf 661}, 263 (2008)
  [hep-ph/0605193].
 
\bibitem{HVTrigger} 
  G.~Aad {\it et al.}  [ATLAS Collaboration],
  ATL-PHYS-PUB-2009-082.

\bibitem{NMSSM} 
  S.~Chang, R.~Dermisek, J.~F.~Gunion and N.~Weiner,
  Ann.\ Rev.\ Nucl.\ Part.\ Sci.\  {\bf 58}, 75 (2008)
  [arXiv:0801.4554 [hep-ph]] and references therein.

 \bibitem{Hidden_Higgs} 
  A.~Falkowski, J.~T.~Ruderman, T.~Volansky and J.~Zupan,
  JHEP {\bf 1005}, 077 (2010)
  [arXiv:1002.2952 [hep-ph]];

  A.~Falkowski, J.~T.~Ruderman, T.~Volansky and J.~Zupan,
  Phys.\ Rev.\ Lett.\  {\bf 105}, 241801 (2010)
  [arXiv:1007.3496 [hep-ph]].

\bibitem{DisplacedSUSY} 
  P.~W.~Graham, D.~E.~Kaplan, S.~Rajendran and P.~Saraswat,
  JHEP {\bf 1207}, 149 (2012)
  [arXiv:1204.6038 [hep-ph]].


\bibitem{DisplacedDM} 
  S.~Chang and M.~A.~Luty,
  arXiv:0906.5013 [hep-ph].
  

\bibitem{SplitSusy} 
  G.~F.~Giudice and A.~Romanino,
  Nucl.\ Phys.\ B {\bf 699}, 65 (2004)
  [Erratum-ibid.\ B {\bf 706}, 65 (2005)]
  [hep-ph/0406088];
  
  N.~Arkani-Hamed, S.~Dimopoulos, G.~F.~Giudice and A.~Romanino,
  Nucl.\ Phys.\ B {\bf 709}, 3 (2005)
  [hep-ph/0409232];
  
  N.~Arkani-Hamed, A.~Gupta, D.~E.~Kaplan, N.~Weiner and T.~Zorawski,
  arXiv:1212.6971 [hep-ph].


  \bibitem{GMSB}  
  S.~Dimopoulos, S.~D.~Thomas and J.~D.~Wells,
  Phys.\ Rev.\ D {\bf 54}, 3283 (1996)
  [hep-ph/9604452];
  
  S.~Dimopoulos, M.~Dine, S.~Raby and S.~D.~Thomas,
  Phys.\ Rev.\ Lett.\  {\bf 76}, 3494 (1996)
  [hep-ph/9601367].
  
  
\bibitem{GMSB:Prompt} 
  P.~Meade, M.~Reece and D.~Shih,
  JHEP {\bf 1005}, 105 (2010)
  [arXiv:0911.4130 [hep-ph]].
 
\bibitem{GMSB:Displaced} 
  P.~Meade, M.~Reece and D.~Shih,
  JHEP {\bf 1010}, 067 (2010)
  [arXiv:1006.4575 [hep-ph]].

\bibitem{NeutralinoOscillation} 
  Y.~Grossman, B.~Shakya and Y.~Tsai,
  arXiv:1211.3121 [hep-ph].

  \bibitem{MassSplittingWino} 
  J.~L.~Feng, T.~Moroi, L.~Randall, M.~Strassler and S.~-f.~Su,
  Phys.\ Rev.\ Lett.\  {\bf 83}, 1731 (1999)
  [hep-ph/9904250].

 
 \bibitem{MDM} 
  M.~Cirelli, N.~Fornengo and A.~Strumia,
  Nucl.\ Phys.\ B {\bf 753}, 178 (2006)
  [hep-ph/0512090].
 
 
 \bibitem{2loopWino} 
  M.~Ibe, S.~Matsumoto and R.~Sato,
  arXiv:1212.5989 [hep-ph].


\bibitem{MG5} 
  J.~Alwall, M.~Herquet, F.~Maltoni, O.~Mattelaer and T.~Stelzer,
  JHEP {\bf 1106}, 128 (2011)
  [arXiv:1106.0522 [hep-ph]].
  
\bibitem{Pythia} 
  T.~Sjostrand, S.~Mrenna and P.~Z.~Skands,
  Comput.\ Phys.\ Commun.\  {\bf 178}, 852 (2008)
  [arXiv:0710.3820 [hep-ph]].

\bibitem{FeynRules} 
  N.~D.~Christensen and C.~Duhr,
  Comput.\ Phys.\ Commun.\  {\bf 180}, 1614 (2009)
  [arXiv:0806.4194 [hep-ph]].


\bibitem{Wh:MSSM} 
  H.~Baer, V.~Barger, A.~Lessa, W.~Sreethawong and X.~Tata,
  Phys.\ Rev.\ D {\bf 85}, 055022 (2012)
  [arXiv:1201.2949 [hep-ph]];
  
  D.~Ghosh, M.~Guchait and D.~Sengupta,
  Eur.\ Phys.\ J.\ C {\bf 72}, 2141 (2012)
  [arXiv:1202.4937 [hep-ph]].

\bibitem{Wh:Prompt} 
  K.~Howe and P.~Saraswat,
  JHEP {\bf 1210}, 065 (2012)
  [arXiv:1208.1542 [hep-ph]].

\bibitem{WWh:fake} 
  M.~Lisanti and N.~Weiner,
  Phys.\ Rev.\ D {\bf 85}, 115005 (2012)
  [arXiv:1112.4834 [hep-ph]];
  
  B.~Feigl, H.~Rzehak and D.~Zeppenfeld,
  Phys.\ Lett.\ B {\bf 717}, 390 (2012)
  [arXiv:1205.3468 [hep-ph]].
  
\bibitem{WWanomaly} 
  D.~Curtin, P.~Jaiswal and P.~Meade,
  arXiv:1206.6888 [hep-ph].
  
  
\bibitem{ATLAS:hWW} 
  G.~Aad {\it et al.}  [ATLAS Collaboration],
  Phys.\ Lett.\ B {\bf 716}, 62 (2012)
  [arXiv:1206.0756 [hep-ex]].

\bibitem{hbb} 
   [ATLAS Collaboration], 
   ATLAS-CONF-2012-161;
   
   [CMS Collaboration],
  CMS-PAS-HIG-12-044.
  





\bibitem{trilepton} 
  G.~Aad {\it et al.}  [ATLAS Collaboration],
  Phys.\ Lett.\ B {\bf 718}, 841 (2013)
  [arXiv:1208.3144 [hep-ex]];
  
  S.~Chatrchyan {\it et al.}  [CMS Collaboration],
  JHEP {\bf 1206}, 169 (2012)
  [arXiv:1204.5341 [hep-ex]];
  
  [ATLAS Collaboration],
  ATLAS-CONF-2012-154.

\bibitem{OSdilepton} 
  G.~Aad {\it et al.}  [ATLAS Collaboration],
  Phys.\ Lett.\ B {\bf 718}, 879 (2013)
  [arXiv:1208.2884 [hep-ex]];

  S.~Chatrchyan {\it et al.}  [CMS Collaboration],
  arXiv:1301.0916 [hep-ex];
  
  [CMS Collaboration],
  CMS-PAS-SUS-12-026.
  
\bibitem{SSOSdilepton} 
  S.~Chatrchyan {\it et al.}  [CMS Collaboration],
  JHEP {\bf 1211}, 147 (2012)
  [arXiv:1209.6620 [hep-ex]];
      
   [CMS Collaboration],
  CMS-PAS-SUS-12-022.
  
\bibitem{SSdilepton} 
  [ATLAS Collaboration],
  ATLAS-CONF-2011-126.
  

\bibitem{WW:SM} 
  S.~Chatrchyan {\it et al.}  [CMS Collaboration],
  arXiv:1301.4698 [hep-ex];
  
  S.~Chatrchyan {\it et al.}  [CMS Collaboration],
  Phys.\ Lett.\ B {\bf 699}, 25 (2011)
  [arXiv:1102.5429 [hep-ex]];
  
  G.~Aad {\it et al.}  [ATLAS Collaboration],
  arXiv:1210.2979 [hep-ex].
 
 
 
\bibitem{LongLivedMain} 
  G.~Aad {\it et al.}  [ATLAS Collaboration],
  arXiv:1211.1597 [hep-ex];
  
  S.~Chatrchyan {\it et al.}  [CMS Collaboration],
  Phys.\ Lett.\ B {\bf 713}, 408 (2012)
  [arXiv:1205.0272 [hep-ex]];

 
 
 
\bibitem{ATLAS:LLChargino} 
  [ATLAS Collaboration],
  JHEP {\bf 1301}, 131 (2013)
  [arXiv:1210.2852 [hep-ex]].
 
\bibitem{LongLived} 
 V.~M.~Abazov {\it et al.}  [D0 Collaboration],
  arXiv:1211.2466 [hep-ex];
  
  T.~Aaltonen {\it et al.}  [CDF Collaboration],
  Phys.\ Rev.\ Lett.\  {\bf 103}, 021802 (2009)
  [arXiv:0902.1266 [hep-ex]].





\bibitem{LLNeutral:GMSB:PhotonMET} 
  [CMS Collaboration],
  CMS-PAS-EXO-11-035;
  
  A.~Abulencia {\it et al.}  [CDF Collaboration],
  Phys.\ Rev.\ Lett.\  {\bf 99}, 121801 (2007)
  [arXiv:0704.0760 [hep-ex]].

\bibitem{LLNeutral:Muon+Jets} 
  [ATLAS Collaboration],
  ATLAS-CONF-2012-113.

\bibitem{LLNeutral:bb} 
  V.~M.~Abazov {\it et al.}  [D0 Collaboration],
  Phys.\ Rev.\ Lett.\  {\bf 103}, 071801 (2009)
  [arXiv:0906.1787 [hep-ex]];
  
  T.~Aaltonen {\it et al.}  [CDF Collaboration],
  Phys.\ Rev.\ D {\bf 85}, 012007 (2012)
  [arXiv:1109.3136 [hep-ex]].
  
\bibitem{LLNeutral:mumu} 
  V.~M.~Abazov {\it et al.}  [D0 Collaboration],
  Phys.\ Rev.\ Lett.\  {\bf 97}, 161802 (2006)
  [hep-ex/0607028].
  
\bibitem{LLNeutral:dilepton} 
    S.~Chatrchyan {\it et al.}  [CMS Collaboration],
  arXiv:1211.2472 [hep-ex].
  
\bibitem{HV:MuonTracker} 
  G.~Aad {\it et al.}  [ATLAS Collaboration],
  Phys.\ Rev.\ Lett.\  {\bf 108}, 251801 (2012)
  [arXiv:1203.1303 [hep-ex]].

\bibitem{LLNeutral:muonjets} 
  G.~Aad {\it et al.}  [ATLAS Collaboration],
  arXiv:1210.0435 [hep-ex];
  
  V.~M.~Abazov {\it et al.}  [D0 Collaboration],
  Phys.\ Rev.\ Lett.\  {\bf 103}, 081802 (2009)
  [arXiv:0905.1478 [hep-ex]];
  
  V.~M.~Abazov {\it et al.}  [D0 Collaboration],
  Phys.\ Rev.\ Lett.\  {\bf 105}, 211802 (2010)
  [arXiv:1008.3356 [hep-ex]].

\bibitem{LLNeutral:diphoton} 
  V.~M.~Abazov {\it et al.}  [D0 Collaboration],
  Phys.\ Rev.\ Lett.\  {\bf 101}, 111802 (2008)
  [arXiv:0806.2223 [hep-ex]].



\bibitem{stopped} 
  S.~Chatrchyan {\it et al.}  [CMS Collaboration],
  JHEP {\bf 1208}, 026 (2012)
  [arXiv:1207.0106 [hep-ex]];

\bibitem{Aad:2012zn} 
  G.~Aad {\it et al.}  [ATLAS Collaboration],
  Eur.\ Phys.\ J.\ C {\bf 72}, 1965 (2012)
  [arXiv:1201.5595 [hep-ex]].


\bibitem{DM:jet} 
  G.~Aad {\it et al.}  [ATLAS Collaboration],
  arXiv:1210.4491 [hep-ex];

  S.~Chatrchyan {\it et al.}  [CMS Collaboration],
  JHEP {\bf 1209}, 094 (2012)
  [arXiv:1206.5663 [hep-ex]].


\bibitem{DM:photon} 
  G.~Aad {\it et al.}  [ATLAS Collaboration],
  arXiv:1209.4625 [hep-ex];

  S.~Chatrchyan {\it et al.}  [CMS Collaboration],
  Phys.\ Rev.\ Lett.\  {\bf 108}, 261803 (2012)
  [arXiv:1204.0821 [hep-ex]].



\bibitem{DM:WZ} 
  Y.~Bai and T.~M.~P.~Tait,
  arXiv:1208.4361 [hep-ph];

  S.~Chatrchyan {\it et al.}  [CMS Collaboration],
  JHEP {\bf 1208}, 023 (2012)
  [arXiv:1204.4764 [hep-ex]];


  L.~M.~Carpenter, A.~Nelson, C.~Shimmin, T.~M.~P.~Tait and D.~Whiteson,
  arXiv:1212.3352 [hep-ex];

  G.~Aad {\it et al.}  [ATLAS Collaboration],
  arXiv:1211.6096 [hep-ex].



\bibitem{b-tagging} 
  [ATLAS Collaboration],
ATLAS-CONF-2011-102;

  S.~Chatrchyan {\it et al.}  [CMS Collaboration],
 arXiv:1211.4462 [hep-ex].


\bibitem{Djouadi2} 
  A.~Djouadi,
  Phys.\ Rept.\  {\bf 459}, 1 (2008)
  [hep-ph/0503173].

  
\end{thebibliography}
\end{document}